\documentclass[english,usenatbib,onecolumn]{mnras}
\usepackage[T1]{fontenc}
\usepackage[latin9]{inputenc}
\usepackage{geometry}
\usepackage{amsmath}
\usepackage{amssymb}
\usepackage{graphicx}
\usepackage{esint}

\makeatletter

\providecommand{\tabularnewline}{\\}

\numberwithin{equation}{section}

\makeatother

\usepackage{babel}
\begin{document}

\title{Large Covariance Matrices: Accurate Models Without Mocks}

\author[O'Connell et al.]{Ross O'Connell$^1$, Daniel J. Eisenstein$^2$
\\
$^{1}$McWilliams Center for Cosmology, Carnegie Mellon University, 5000 Forbes Ave, Pittsburgh, PA 15213, USA\\
$^{2}$Harvard-Smithsonian Center for Astrophysics, 60 Garden St., Cambridge, MA 02138, USA
}
\maketitle
\begin{abstract}
Covariance matrix estimation is a persistent challenge for cosmology.
We focus on a class of model covariance matrices that can be generated
with high accuracy and precision, using a tiny fraction of the computational
resources that would be required to achieve comparably precise covariance matrices using
mock catalogues. In previous work, the free parameters in these models
were determined using sample covariance matrices computed using
a large number of mocks, but we demonstrate that those parameters
can be estimated consistently and with good precision by applying
jackknife methods to a single survey volume. This enables model covariance
matrices that are calibrated from data alone, with no reference to
mocks.
\end{abstract}

\section{Introduction}

No matter the scope of a cosmological survey, we have only one sky
to observe. This complicates the statistical analysis of cosmological
surveys. A common approach is to generate a large number of independent,
synthetic skies, then apply standard sample statistics to them. The
readily apparent limitations of this approach are that it is challenging
to ensure that the synthetic skies reflect the physics of the actual
universe, and that the computational cost of generating these ``mock
catalogues'' can be substantial. In this paper we take an approach
introduced in \cite{OConnell:2015src}, which generates the covariance
matrix for a galaxy correlation function with correct long-distance
physics and survey geometry, and extend it so that the \emph{short-distance}
physics can be calibrated directly against a survey, without reference
to mocks.

Using mock catalogues to generate a covariance matrix requires a large number of reasonably accurate mocks. The consequences
of having an insufficient number of mocks have received significant
attention. If $n_{\text{samples}}$ mocks are used to generate a covariance
matrix for a correlation function estimated using $n_{\text{bins}}$
different bins the scale for noise in the covariance matrix is set
by $n_{\text{bins}}/n_{\text{samples}}$. Noise in the covariance
matrix propagates through to become additional noise on cosmological
parameter estimates, increasing the parameter covariance by a factor
of $n_{\text{bins}}/n_{\text{samples}}$\textbf{ }\citep{Dodelson:2013uaa,Percival:2013sga}.
We note that the next generation of surveys, including Euclid \citep{EuclidRedBook}, 
DESI \citep{Levi:2013gra},
and WFIRST \citep{Dore:2018smn} aim at tomographic analyses
 and that a simultaneous analysis
of multiple redshift bins will dramatically increase the number of
correlation function bins used.

In \cite{OConnell:2015src} we developed a method built on the observation,
due to Bernstein \citep{Bernstein1994}, that the covariance matrix
of a 2-point correlation function can itself by written in terms of
correlations between four points, integrated over the survey volume.
We perform these integrals using a realistic 2-point correlation
function and accurately representation of the survey geometry, to produce
a covariance matrix that accurately reflects the long-distance physics
and structure of the survey. Related work includes \cite{Pearson:2015gca},
which constructed a simple model of the power spectrum covariance
matrix, and \cite{Grieb:2015bia}, which took a similar approach
to that in\textbf{ }\cite{OConnell:2015src} but on a cubic, uniform
survey. 

\begin{figure}
\includegraphics[width=\textwidth]{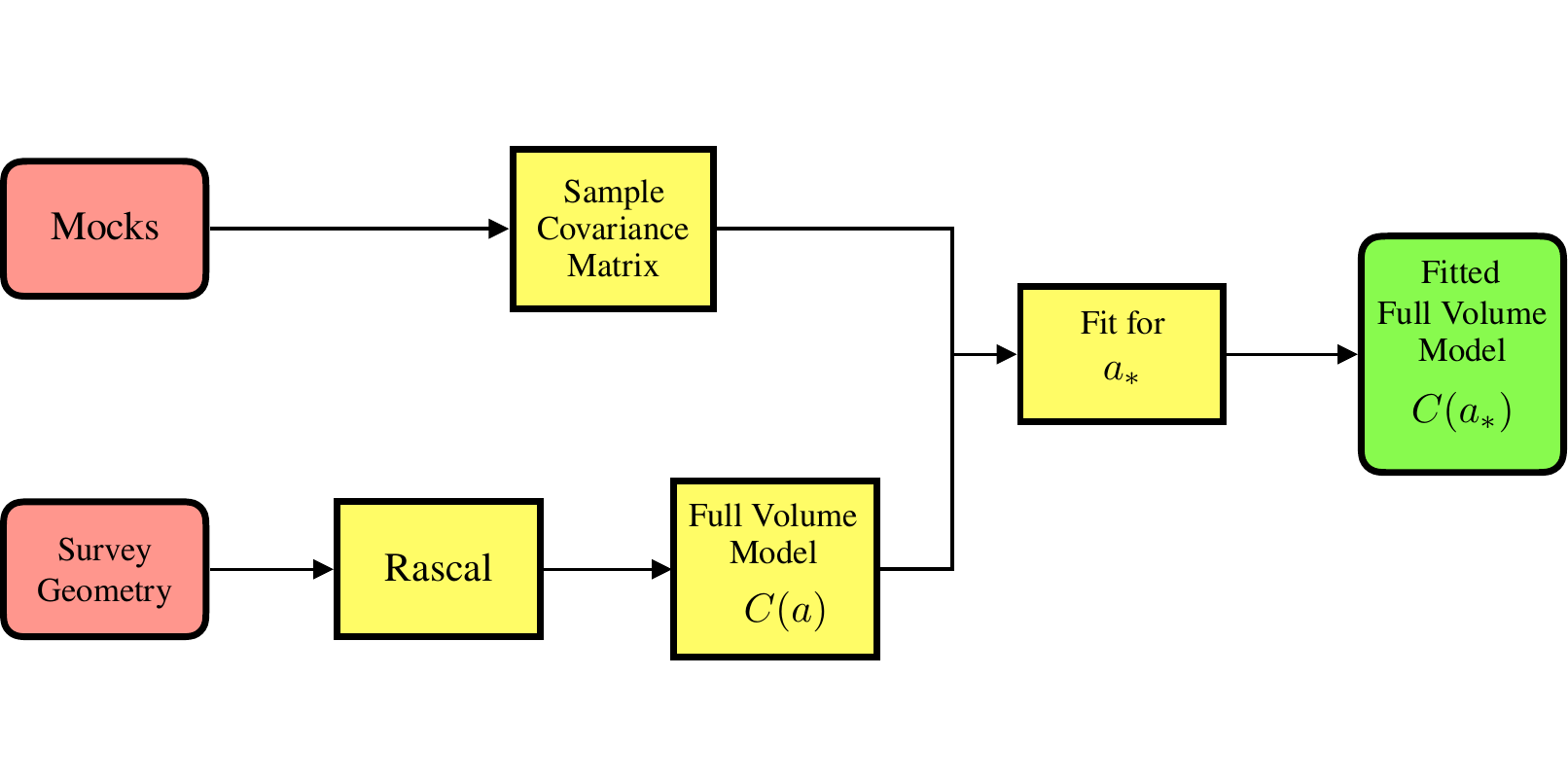}

\caption{\label{fig:OldFlowChart}The procedure introduced in \citep{OConnell:2015src}
for generating and calibrating a model covariance matrix. The model
includes one unknown parameter, $a$, the shot-noise rescaling, which
is calibrated using mock catalogues.}

\end{figure}

The four-point correlations noted above include contributions from
the connected 3- and 4-point galaxy correlation functions, which are
not as well-understood as the 2-point function. \cite{OConnell:2015src} approximated these contributions by  introducing
a shot-noise rescaling parameter, $a$, which effectively modelled
the short-distance contributions to the covariance matrix from the
3- and 4-point functions as an increase in shot-noise. 
$a$ was estimated using mock catalogues and the resulting fit covariance
matrix was found to be both accurate and precise. Critically, this required a tiny fraction
of the computational time required to generate a mock covariance matrix
of comparable precision. The procedure is illustrated in figure \ref{fig:OldFlowChart}.
This approach was further tested in \cite{Vargas-2016}, where it
was found that the resulting covariance matrix performed at least
as well as a mock covariance matrix for BAO measurements with BOSS
\citep{2013AJ....145...10D}.

\begin{figure}
\includegraphics[width=\textwidth]{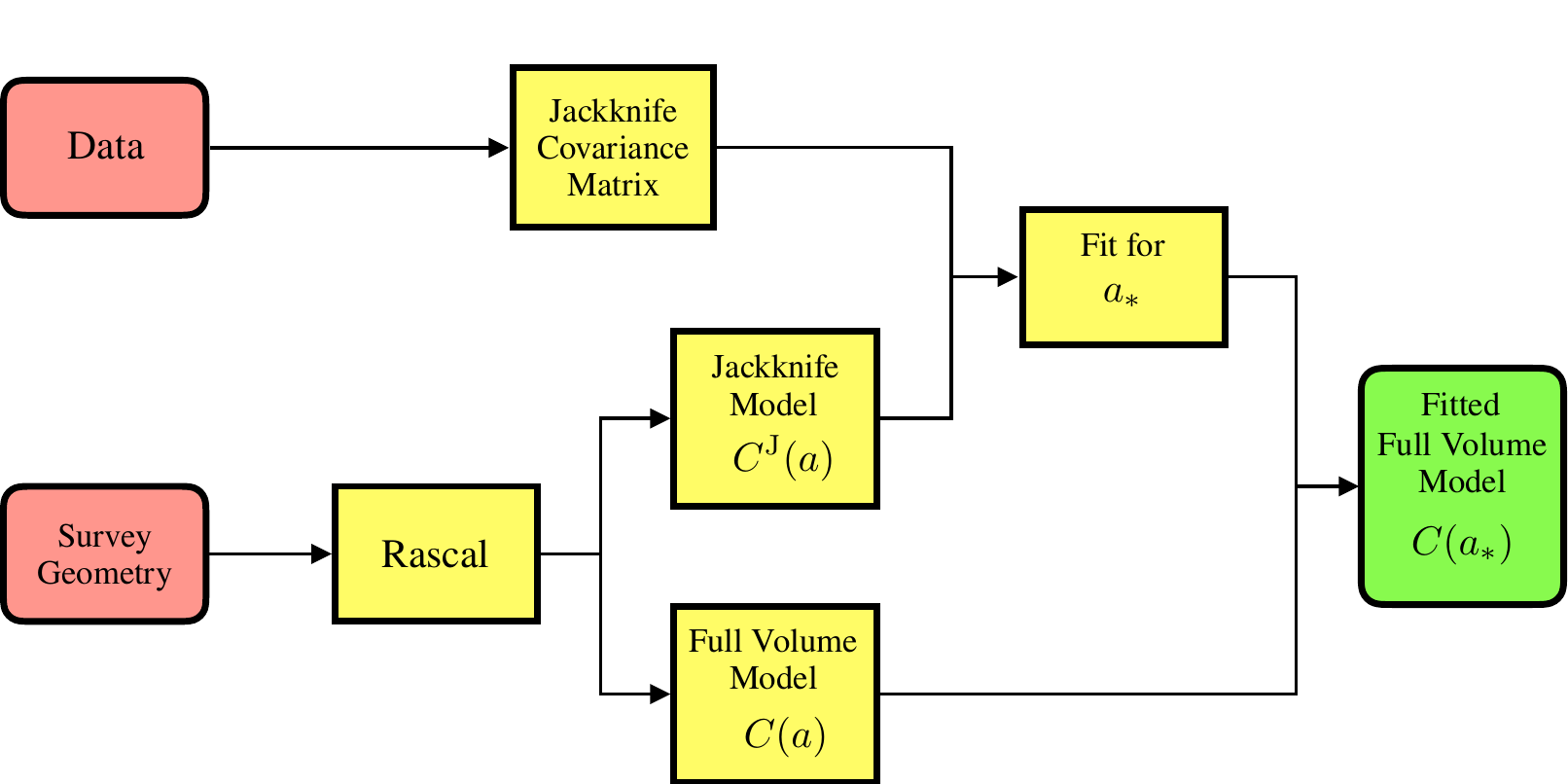}

\caption{\label{fig:NewFlowChart}The procedure proposed in this paper for
generating and calibrating a model covariance matrix. The unknown
parameter $a$ is calibrated using the data directly, rather than
mock catalogues. This is accomplished using jackknife methods.}
\end{figure}

In this paper, we propose to estimate the shot-noise rescaling $a$
using actual survey data, rather than mock catalogues. The essential
observation is that since $a$ is being used to model short-distance
physics, we need not use or mimic the entire survey volume in order
to estimate it. Instead we propose to use the actual survey to generate
a jackknife covariance matrix, then use the jackknife covariance
matrix to estimate $a$. 
The computational cost to do this is quite low, since generating the jackknife covariance matrix requires very little computation beyond counting the pairs in the survey. 
The estimated value of $a$ can then be
used in the original model covariance matrix. This new procedure is
illustrated in figure \ref{fig:NewFlowChart}. We find
that the level of precision on $a$ that can be achieved with a single
survey volume is ample for many applications. It is therefore possible
to perform covariance matrix estimation for upcoming surveys without reference
to mock catalogues.

The observation that relatively small volumes can provide usable information
about the covariance matrix has generated recent interest. 
 \cite{Klypin:2017} investigated the power spectrum covariance
matrix using small-volume cubic mocks. Small-volume
cubic simulations were also used in\textbf{ }\cite{Howlett:2017vwp}
to generate a scaled covariance matrix for the 2-point correlation
function. The jackknife approach introduced here allows us to utilise
small-scale information while accurately reflecting the true survey
geometry.

In light of the urgency of the covariance matrix problem for upcoming
surveys, many approaches to the problem are currently being developed:
\begin{itemize}
\item New techniques in mock generation aim to increase $n_{\text{samples}}$.
For overviews of recent progress see \cite{Chuang:2014toa} and \cite{Lippich:2018wrx}.
\item Compression of the correlation function can reduce $n_{\text{bins}}$.
This can be particularly helpful in analysing tomographic data. A
prominent example is the ``redshift weights'' approach\textbf{ }introduced
in \cite{Zhu:2014ica} and most recently applied in \cite{Zhu:2018edv}.
\item Several empirical techniques have been developed to smooth sample
covariance matrices computed from mocks. These take advantage of resampling
methods \citep{Escoffier:2016qnf}, shrinkage \citep{Joachimi:2016xhk},
or the sparse structure of the precision matrix (the inverse of the
covariance matrix) \citep{Padmanabhan:2015vlf}.
\end{itemize}
The result is that practitioners can combine a variety of physical
and statistical insights when analysing a cosmological survey. We
hope that our contribution will be useful in this regard.

This paper is organised as follows. In section \ref{sec:Review} we
briefly review the results of \cite{OConnell:2015src}, including
the full and Gaussian model covariance matrices and the 1-parameter
model for non-Gaussian contributions. In section \ref{sec:Jackknife}
we specify how we will compute a jackknife covariance matrix from
a single survey volume and how we will compute the corresponding
jackknife \emph{model} covariance matrix. In section \ref{sec:Validation-Methods}
we use mocks to verify that the values of $a$ estimated from single
survey volumes, using a jackknife, are consistent with the values
that would be estimated from those mocks using a sample covariance.
This establishes the consistency of our method and provides evidence
for our claim that $a$ is modelling short-distance physics. We conclude
in section \ref{sec:Outlook}. In appendix \ref{sec:Inversion} we
present a jackknife-inspired method for accurately inverting a model
covariance matrix. That method is used in this paper and may be of
interest to researchers working on model covariance matrices in other
contexts.

\section{Model Covariance Matrices}
\label{sec:Review}

Given an estimator $\hat{\xi}_{a}$ for a correlation function, the
covariance matrix for that estimator is
\begin{equation}
C_{ab}=\left\langle \hat{\xi}_{a}\hat{\xi}_{b}\right\rangle -\left\langle \hat{\xi}_{a}\right\rangle \left\langle \hat{\xi}_{b}\right\rangle .
\end{equation}
In practice $C_{ab}$ itself is often estimated by evaluating $\hat{\xi}_{a}$
on a large number of mock catalogues, then computing a sample covariance
$\hat{C}_{ab}$ from the $\hat{\xi}_{a}$. The idea of a model covariance
matrix is to use theoretical insights to produce a more direct estimate
of $C_{ab}$. The primary elements of the model we use are the shot
noise (which may vary across the survey) and the 2-point correlation
function, both of which are assumed to be well-understood. The benefit
of the model covariance approach is that high precision for the covariance
matrix is readily attained, while the mock approach requires a significant
investment of computational resources to achieve even modest degrees
of precision. In \cite{Percival:2013sga} it was shown that insufficient
precision in the covariance matrix for the correlation function propagates
through to reduce the precision of cosmological measurements performed
with the correlation function, so methods to improve the precision
of covariance matrices have immediate value for observational cosmologists.

In the model covariance matrix approach a primary challenge is finding
an accurate way to approximate non-Gaussian contributions to $C_{ab}$.
These are determined by the 3- and 4-point correlation functions,
which in most applications are only partially understood. Many approaches
to the non-Gaussian contributions are possible; a simple model was
introduced in \cite{OConnell:2015src} to approximate these contributions
by rescaling the shot noise in the survey by a uniform factor $a$.
In this section we briefly review the results of \cite{OConnell:2015src},
including the full and Gaussian model covariance matrices and the
1-parameter model for non-Gaussian contributions. 

\subsection{Covariance Matrix from $n$-point Functions}
\label{subsec:Review-basics}

To illustrate the method we consider the 2-point correlation function
in a galaxy survey. We begin by breaking the survey into a large number
of non-overlapping cells, such that each cell contains either one
or zero galaxies. Let $d_{i}$ be the number of galaxies in cell $i$.
We also introduce $n_{i}$, the number density of galaxies in cell
$i$, and $w_{i}$, the weight applied to cell $i$, to account for
possible inhomogeneities in the survey. The overdensity in cell $i$
is then 
\begin{equation}
\delta_{i}=\frac{d_{i}}{n_{i}}-1.
\end{equation}
The estimate of the correlation function in bin $a$ is 
\begin{align}
\hat{\xi}_{a} & =\frac{1}{\text{RR}_{a}}\sum_{i\neq j}\Theta_{a}^{ij}w_{i}w_{j}\delta_{i}\delta_{j}\,,\\
\text{RR}_{a} & =\sum_{i\neq j}\Theta_{a}^{ij}w_{i}w_{j}\delta_{i}\delta_{j}\,,
\end{align}
where $\Theta_{a}^{ij}$ is a binning matrix that is one when the
separation between cells $i$ and $j$ falls in correlation function
bin $a$ and zero otherwise. In the following we will assume that
the binning matrices are symmetric, $\Theta_{a}^{ij}=\Theta_{a}^{ji}$,
as is appropriate when estimating an autocorrelation function.

The covariance matrix for $\hat{\xi}_{a}$ is 
\begin{align}
C_{ab} & =\left\langle \hat{\xi}_{a}\hat{\xi}_{b}\right\rangle -\left\langle \hat{\xi}_{a}\right\rangle \left\langle \hat{\xi}_{b}\right\rangle \\
 & =\frac{1}{\text{RR}_{a}\text{RR}_{b}}\sum_{i\neq j}\sum_{k\neq\ell}\Theta_{a}^{ij}\Theta_{b}^{k\ell}n_{i}n_{j}n_{k}n_{\ell}w_{i}w_{j}w_{k}w_{\ell}\left[\left\langle \delta_{i}\delta_{j}\delta_{k}\delta_{\ell}\right\rangle -\left\langle \delta_{i}\delta_{j}\right\rangle \left\langle \delta_{k}\delta_{\ell}\right\rangle \right].
\end{align}
In order to connect this expression to $n-$point functions we need
to remove contributions to the sum from overlapping cells, e.g.\ where
$i=k$. Such contributions are readily simplified by the following
identity,
\begin{equation}
\delta_{i}^{2}\approx\frac{1}{n_{i}}\left(1+\delta_{i}\right),
\end{equation}
which we think of as a contraction between two $\delta$'s. Performing
the required contractions and exploiting the symmetry of the binning
matrices $\Theta_{a}^{ij}$, we find 
\begin{align}
C_{ab}= & \frac{1}{\text{RR}_{a}\text{RR}_{b}}\left[\sum_{i\neq j\neq k\neq\ell}n_{i}n_{j}n_{k}n_{\ell}w_{i}w_{j}w_{k}w_{\ell}\Theta_{a}^{ij}\Theta_{b}^{k\ell}\left[\left\langle \delta_{i}\delta_{j}\delta_{k}\delta_{\ell}\right\rangle -\left\langle \delta_{i}\delta_{j}\right\rangle \left\langle \delta_{k}\delta_{\ell}\right\rangle \right]\right.\nonumber \\
 & \hphantom{\frac{n^{4}}{\text{RR}_{a}\text{RR}_{b}}}+4\sum_{i\neq j\neq k}n_{i}n_{j}n_{k}w_{i}^{2}w_{j}w_{k}\Theta_{a}^{ij}\Theta_{b}^{ki}\left\langle \left(1+\delta_{i}\right)\delta_{j}\delta_{k}\right\rangle \nonumber \\
 & \left.\hphantom{\frac{n^{4}}{\text{RR}_{a}\text{RR}_{b}}}+2\delta_{ab}\sum_{i\neq j}n_{i}n_{j}w_{i}^{2}w_{j}^{2}\Theta_{a}^{ij}\left\langle \left(1+\delta_{i}\right)\left(1+\delta_{j}\right)\right\rangle \right].\label{eq:Standard_Cov}
\end{align}
The connection to $n-$point correlation functions is now straightforward,
\begin{align}
C_{ab}= & \frac{1}{\text{RR}_{a}\text{RR}_{b}}\left[\sum_{i\neq j\neq k\neq\ell}n_{i}n_{j}n_{k}n_{\ell}w_{i}w_{j}w_{k}w_{\ell}\Theta_{a}^{ij}\Theta_{b}^{k\ell}\left(\xi_{ik}^{\left(2\right)}\xi_{j\ell}^{\left(2\right)}+\xi_{i\ell}^{\left(2\right)}\xi_{jk}^{\left(2\right)}+\xi_{ijkl}^{\left(4\right)}\right)\right.\nonumber \\
 & \hphantom{\frac{n^{4}}{\text{RR}_{a}\text{RR}_{b}}}+4\sum_{i\neq j\neq k}n_{i}n_{j}n_{k}w_{i}^{2}w_{j}w_{k}\Theta_{a}^{ij}\Theta_{b}^{ki}\left(\xi_{jk}^{\left(2\right)}+\xi_{ijk}^{\left(3\right)}\right)\nonumber \\
 & \left.\hphantom{\frac{n^{4}}{\text{RR}_{a}\text{RR}_{b}}}+2\delta_{ab}\sum_{i\neq j}n_{i}n_{j}w_{i}^{2}w_{j}^{2}\Theta_{a}^{ij}\left(1+\xi_{ij}^{\left(2\right)}\right)\right],\label{eq:Standard_Cov_Corr}
\end{align}
where $\xi_{ij}^{\left(2\right)}$ is the familiar 2-point correlation
function, $\xi_{ijk}^{\left(3\right)}$ is the 3-point correlation
function, and $\xi_{ijk\ell}^{\left(4\right)}$ is the 4-point correlation
function. A continuum limit yields the expressions familiar from \cite{Bernstein1994}
and \cite{OConnell:2015src}.

\subsection{Modelling Non-Gaussianity}
\label{subsec:Review-Non-Gaussianity}

If we look over (\ref{eq:Standard_Cov_Corr}), we see that it includes
$n_{i}$ and $w_{i}$, which as survey properties are assumed to be
readily available. The 2-point function $\xi_{ij}^{\left(2\right)}$
also appears, and in most applications is understood with sufficient
accuracy to compute its contribution to $C_{ab}$. The 3- and 4-point
functions $\xi_{ijk}^{\left(3\right)}$ and $\xi_{ijk\ell}^{\left(4\right)}$,
on the other hand, are often only partially understood. When an accurate
version of the 3- and/or 4-point functions is not available we could
consider a variety of models for non-Gaussian contributions to $C_{ab}$.
One such model approximates the contributions of the 3- and 4-point
functions, which we expect to be most relevant at small separations,
as additional shot noise. We do this by splitting up the Gaussian
contributions in (\ref{eq:Standard_Cov_Corr}) as follows:
\begin{align}
C_{4,ab} & =\frac{1}{\text{RR}_{a}\text{RR}_{b}}\sum_{i\neq j\neq k\neq\ell}n_{i}n_{j}n_{k}n_{\ell}w_{i}w_{j}w_{k}w_{\ell}\Theta_{a}^{ij}\Theta_{b}^{k\ell}\left(\xi_{ik}^{\left(2\right)}\xi_{j\ell}^{\left(2\right)}+\xi_{i\ell}^{\left(2\right)}\xi_{jk}^{\left(2\right)}\right),\label{eq:C4_Full}\\
C_{3,ab} & =\frac{4}{\text{RR}_{a}\text{RR}_{b}}\sum_{i\neq j\neq k}n_{i}n_{j}n_{k}w_{i}^{2}w_{j}w_{k}\Theta_{a}^{ij}\Theta_{b}^{ki}\xi_{jk}^{\left(2\right)}\,,\label{eq:C3_Full}\\
C_{2,ab} & =\frac{2}{\text{RR}_{a}\text{RR}_{b}}\delta_{ab}\sum_{i\neq j}n_{i}n_{j}w_{i}^{2}w_{j}^{2}\Theta_{a}^{ij}\left(1+\xi_{ij}^{\left(2\right)}\right)\,.\label{eq:C2_Full}
\end{align}
We can then implement a uniform increase in the shot noise by a factor
$a$ by taking $n_{i}\to n_{i}/a$. Recall that $\text{RR}_{a}\propto n^{-2}$,
so the non-Gaussian model is 
\begin{equation}
C_{ab}\left(a\right)=C_{4,ab}+aC_{3,ab}+a^{2}C_{2,ab}\,.\label{eq:NG_Model_Full}
\end{equation}
In \cite{OConnell:2015src} it was shown that the unknown parameter
$a$ can be determined by fitting\footnote{An updated discussion of fitting methods can be found in section \ref{sec:Validation-Methods}.}
the model to the sample covariance computed from mock catalogues. The
resulting model covariance matrix was found to provide suitable accuracy
for BOSS-like surveys. Numerical integration techniques introduced
in \cite{OConnell:2015src} yield precision that dramatically outstrips
what can currently be achieved with mocks and require only very modest
computational resources ($\approx1,000$ CPU hours). This combination
of accuracy and precision makes the model covariance matrix approach
appealing for future studies of large scale structure and motivates
our further development of the method here.

\section{Jackknife Methods and Model Covariance Matrices}
\label{sec:Jackknife}

One of the limitations of the method described in section \ref{subsec:Review-Non-Gaussianity}
is that it relies on mock catalogues to calibrate the unknown parameter
$a$, which in turn ensures the accuracy of the model covariance matrix.
While fewer mocks are required to calibrate $a$ than are required
to generate a precise sample covariance matrix, the potential accuracy
of the model covariance matrix is limited in part by the accuracy
of the mock catalogues. For example, if the connected 4-point function
in the mocks is smaller than the connected 4-point function in the
actual survey, we expect this to result in a biased estimate of $a$.

Fortunately any reasonably large survey includes sufficient volume
to make multiple estimates of the correlation function, and thus contains
information about the statistics of the correlation function. One
could consider a variety of techniques to extract this information,
but here we will use a simple resampling technique, the jackknife,
to make an estimate of $a$ from a single survey volume. We speculate
that in some cases the estimate for $a$ obtained from the actual
survey data will be sufficient for analysis, making mock catalogues
necessary only for controlling biases in the recovered parameters due to systematic effects. When the survey
itself does not yield a sufficiently precise estimate of $a$, we
can still make improvements by applying jackknife methods to each
mock in turn, then combining the results to get a more precise estimate
of $a$ than the sample covariance of those mocks would allow. 

We emphasise that, from the point of view of computational costs, these improvements in the precision of $a$ are essentially free. The jackknife procedure that we will describe requires that the pairs in each survey volume be counted only once, and the only change in computational requirements, relative to standard pair-counting, is that separate counts are maintained for each jackknife region. In other words, the computational time required to use our jackknife procedure to estimate $a$ with a single survey volume really is $\mathcal{O}(1\%)$ of the time required to estimate $a$ using a sample covariance computed from 100 mocks.

\subsection{The Restricted Jackknife}
\label{subsec:Jackknife-Restricted}

In the cosmological version of the jackknife a survey is split into
$n_{\text{jack}}$ regions, then the analysis is repeated $n_{\text{jack}}$
times, with a different region \emph{left out} of the analysis each
time. The results can then be combined to provide an estimate of uncertainties
associated with the analysis. There are several issues that make cosmological
jackknives more complicated to analyse than the traditional statistical
jackknife: 
\begin{enumerate}
\item In analyses that utilise pair counts, some pairs will straddle two
jackknife regions.
\item The regions will generally have different shapes and/or areas.
\item Different regions of a cosmological survey are \emph{not} statistically
independent of one another, as is assumed for the traditional jackknife.
\end{enumerate}
Each of these issues will be relevant in our analysis. 

For the ``restricted'' jackknife, we will simply \emph{exclude any
pairs that straddle two jackknife regions}. We choose this for the
sake of simplicity and anticipate that a jackknife with a more careful
treatment of those pairs could be used without difficulty. The estimate
of the correlation function in a single region $A$ is
\begin{align}
\hat{\xi}_{aA} & =\frac{1}{\text{RR}_{aA}}\sum_{i\neq j}q_{Ai}q_{Aj}\Theta_{a}^{ij}n_{i}n_{j}w_{i}w_{j}\delta_{i}\delta_{j}\,,\\
\text{RR}_{aA} & =\sum_{i\neq j}q_{Ai}q_{Aj}\Theta_{a}^{ij}n_{i}n_{j}w_{i}w_{j}\,.
\end{align}
Jackknife regions are specified by $q_{Ai}$, with $q_{Ai}=1$ if
cell $i$ is in jackknife region $A$, and $q_{Ai}=0$ otherwise.\textbf{ }

We do not assume that the jackknife regions all have the same volume.
This is done for two reasons. First, a greater variety of methods
for generating jackknife regions can be used if we do not require
that all regions have the same volume. In particular, many simple
methods, which are more clearly specified and easily communicated
to other researchers, do not lead to regions with the same volume.
Second, for analyses of the 2-point correlation function the most
appropriate notion of ``volume'' is $\text{RR}_{aA}$, which depends
on the bin being considered. While we can imagine adaptive techniques
to match jackknife region \emph{volumes}, it will not in general be
possible to match $\text{RR}_{aA}$ across all regions $A$ \emph{and}
for all bins $a$.

\begin{figure}
\includegraphics[width=3.25in]{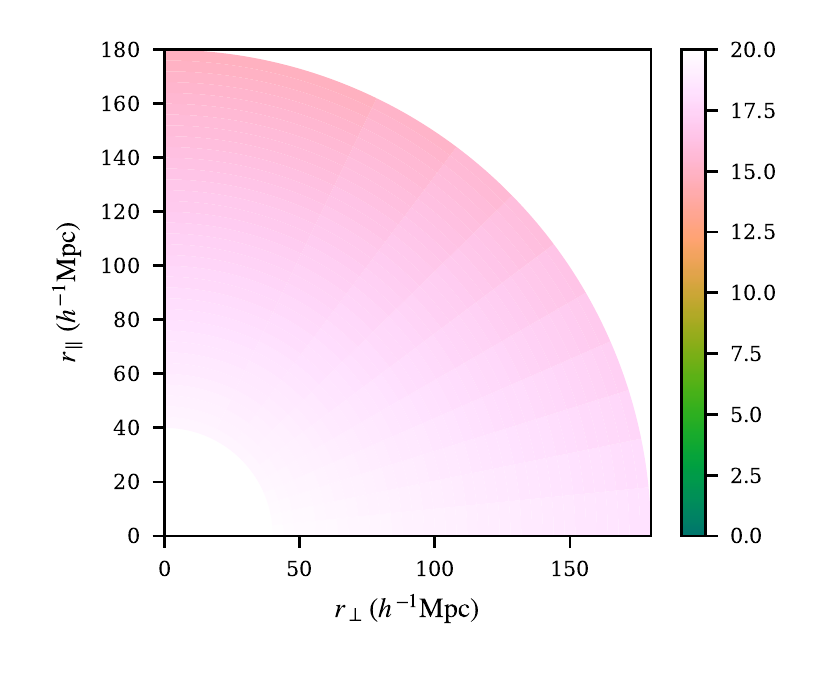}\hfill{}\includegraphics[width=3.25in]{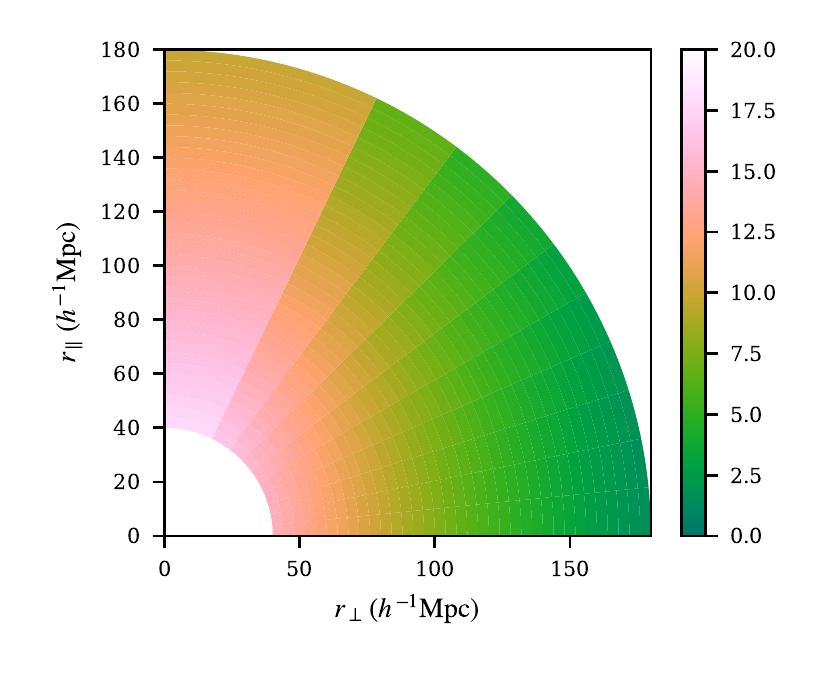}

\caption{\label{fig:RR}Plots of $\text{RR}_{a}/r_{a}^{2}$ for the full survey
(left) and $\text{RR}_{a}^{\text{J}}/r_{a}^{2}$ for the restricted
jackknife (right). Dividing by $r_{a}^{2}$ removes the leading scaling
for $\text{RR}_{a}$. We use jackknife regions that are defined in
angular coordinates (for more details see section \ref{subsec:validation-Jackknife}).
Pairs transverse to the line-of-sight are more likely to cross a boundary
between jackknife regions, and thus be excluded. The restricted jackknife
therefore ``sees'' a geometry that is quite different from the full
survey.}

\end{figure}

To combine the different jackknife regions into a single estimate
we introduce the following weights,
\begin{align}
w_{aA} & =\frac{\text{RR}_{aA}}{\text{RR}_{a}^{\text{J}}}\,,\label{eq:Region_Weights}\\
\text{RR}_{a}^{\text{J}} & =\sum_{A}\text{RR}_{aA}\,.
\end{align}
Note that the relative weighting between regions depends on which
correlation function bin $a$ we consider. The analogue of the full
correlation function in this approach is a weighted sum of the estimates
in each region:
\begin{align}
\hat{\xi}_{a}^{\text{J}} & =\sum_{A}w_{aA}\hat{\xi}_{aA}\\
 & =\frac{1}{\text{RR}_{a}^{\text{J}}}\sum_{A}\sum_{i\neq j}q_{Ai}q_{Aj}\Theta_{a}^{ij}n_{i}n_{j}w_{i}w_{j}\delta_{i}\delta_{j}\\
 & =\frac{1}{\text{RR}_{a}^{\text{J}}}\sum_{i\neq j}Q_{ij}\Theta_{a}^{ij}n_{i}n_{j}w_{i}w_{j}\delta_{i}\delta_{j}\,,\\
Q_{ij} & =\sum_{A}q_{Ai}q_{Aj}.
\end{align}
Note that $Q_{ij}=1$ if cells $i$ and $j$ fall in the same jackknife
region and that $Q_{ij}=0$ otherwise. We emphasise that $\hat{\xi}_{a}^{\text{J}}$
is \emph{not} equivalent to $\hat{\xi}_{a}$, as the inclusion of
$Q_{ij}$ means that pairs that straddle two jackknife regions make
no contribution to $\hat{\xi}_{a}$. A simple way to visualise the
difference between the two approaches is to plot $\text{RR}_{a}$
and $\text{RR}_{a}^{\text{J}}$, as we have in Figure \ref{fig:RR}.
The geometry of the jackknife regions is described in detail in section
\ref{subsec:validation-Jackknife}, but for now it is sufficient to
note that our choice of jackknife regions cuts off many pairs at large
transverse separation, and this is reflected in $\text{RR}_{a}^{\text{J}}$.

We use the standard formulation of a weighted jackknife covariance:
\begin{align}
\hat{C}_{ab}^{\text{J}} & =\frac{1}{1-\sum_{B}w_{aB}w_{bB}}\left[\sum_{A}w_{aA}w_{bA}\left(\hat{\xi}_{aA}-\hat{\xi}_{a}^{\text{J}}\right)\left(\hat{\xi}_{bA}-\hat{\xi}_{b}^{\text{J}}\right)\right]\label{eq:Jackknife_C}
\end{align}
Because $\hat{\xi}_{a}^{J}$ can be written as a weighted average
of $\hat{\xi}_{aA}$, the usual prescription in terms of dropped regions
reduces to a rescaled sample covariance. The standard formulation
of the jackknife covariance provides an unbiased estimate of the true
covariance \emph{when the jackknife regions are independent}. Clearly
that's not the case here, but we'll find that it is still a reasonable
choice. 

\subsection{Model Covariance for the Restricted Jackknife}
\label{subsec:Jackknife-Model}

As noted above, the jackknife covariance and standard covariance ``see''
quite different survey geometries, and so it would not be appropriate
to fit $C\left(a\right)$ to $\hat{C}^{\text{J}}$. For this reason
we will compute a new model, $C^{\text{J}}$, and compare that to
$\hat{C}^{\text{J}}$. The covariance of $\hat{\xi}^{\text{J}}$ is
straightforward to compute: 
\begin{align}
C_{ab}^{\text{J}} & =\left\langle \hat{\xi}_{a}^{\text{J}}\hat{\xi}_{b}^{\text{J}}\right\rangle -\left\langle \hat{\xi}_{a}^{\text{J}}\right\rangle \left\langle \hat{\xi}_{b}^{\text{J}}\right\rangle \\
 & =\frac{1}{\text{RR}_{a}^{\text{J}}\text{RR}_{b}^{\text{J}}}\sum_{i\neq j}\sum_{k\neq\ell}Q_{ij}Q_{k\ell}\Theta_{a}^{ij}\Theta_{b}^{k\ell}n_{i}n_{j}w_{i}w_{j}\left[\left\langle \delta_{i}\delta_{j}\delta_{k}\delta_{\ell}\right\rangle -\left\langle \delta_{i}\delta_{j}\right\rangle \left\langle \delta_{k}\delta_{\ell}\right\rangle \right].
\end{align}
As in the previous section, we must perform a series of contractions
in order to arrive at expressions that can be interpreted in terms
of $n$-point functions. The new twist is that we need to be able
to contract the $Q$'s:
\begin{align}
Q_{ij}Q_{ki} & =\sum_{A,B}q_{Ai}q_{Aj}q_{Bk}q_{Bi}\\
 & =\sum_{A,B}\delta_{AB}q_{Ai}q_{Aj}q_{Bk}\\
 & =\sum_{A}q_{Ai}q_{Aj}q_{Ak}\\
 & =Q_{ijk}\,,\\
Q_{ij}Q_{ji} & =\sum_{A,B}q_{Ai}q_{Aj}q_{Bj}q_{Bi}\\
 & =\sum_{A,B}\delta_{AB}q_{Ai}q_{Aj}\\
 & =Q_{ij}\,.
\end{align}
The expression for the covariance of $\hat{\xi}^{\text{J}}$ is then
\begin{align}
C_{ab}^{\text{J}}= & \frac{1}{\text{RR}_{a}^{\text{J}}\text{RR}_{b}^{\text{J}}}\left[\sum_{i\neq j\neq k\neq\ell}Q_{ij}Q_{k\ell}\Theta_{a}^{ij}\Theta_{b}^{k\ell}n_{i}n_{j}n_{k}n_{\ell}w_{i}w_{j}w_{k}w_{\ell}\left[\left\langle \delta_{i}\delta_{j}\delta_{k}\delta_{\ell}\right\rangle -\left\langle \delta_{i}\delta_{j}\right\rangle \left\langle \delta_{k}\delta_{\ell}\right\rangle \right]\right.\nonumber \\
 & \hphantom{\frac{n^{4}}{\text{RR}_{a}\text{RR}_{b}}}+4\sum_{i\neq j\neq k}Q_{ijk}\Theta_{a}^{ij}\Theta_{b}^{ki}n_{i}n_{j}n_{k}w_{i}^{2}w_{j}w_{k}\left\langle \left(1+\delta_{i}\right)\delta_{j}\delta_{k}\right\rangle \nonumber \\
 & \left.\hphantom{\frac{n^{4}}{\text{RR}_{a}\text{RR}_{b}}}+2\delta_{ab}\sum_{i\neq j}Q_{ij}\Theta_{a}^{ij}n_{i}n_{j}w_{i}^{2}w_{j}^{2}\left\langle \left(1+\delta_{i}\right)\left(1+\delta_{j}\right)\right\rangle \right].\label{eq:Rest_Model}
\end{align}
In words, sets of four points only contribute if $i$ and $j$ fall
in a single jackknife region, and $k$ and $\ell$ also fall in a
single jackknife region. Sets of two or three points only contribute
if \emph{all} points are in the same jackknife region. 

As with the original model we can identify three Gaussian contributions
to the jackknife covariance:
\begin{align}
C_{4,ab}^{\text{J}} & =\frac{1}{\text{RR}_{a}^{\text{J}}\text{RR}_{b}^{\text{J}}}\sum_{i\neq j\neq k\neq\ell}n_{i}n_{j}n_{k}n_{\ell}w_{i}w_{j}w_{k}w_{\ell}Q_{ij}Q_{k\ell}\Theta_{a}^{ij}\Theta_{b}^{k\ell}\left(\xi_{ik}^{\left(2\right)}\xi_{j\ell}^{\left(2\right)}+\xi_{i\ell}^{\left(2\right)}\xi_{jk}^{\left(2\right)}\right),\\
C_{3,ab}^{\text{J}} & =\frac{4}{\text{RR}_{a}^{\text{J}}\text{RR}_{b}^{\text{J}}}\sum_{i\neq j\neq k}n_{i}n_{j}n_{k}w_{i}^{2}w_{j}w_{k}Q_{ijk}\Theta_{a}^{ij}\Theta_{b}^{ki}\xi_{jk}^{\left(2\right)}\,,\\
C_{2,ab}^{\text{J}} & =\frac{2}{\text{RR}_{a}^{\text{J}}\text{RR}_{b}^{\text{J}}}\delta_{ab}\sum_{i\neq j}n_{i}n_{j}w_{i}^{2}w_{j}^{2}Q_{ij}\Theta_{a}^{ij}\left(1+\xi_{ij}^{\left(2\right)}\right)\,.
\end{align}
We use (\ref{eq:NG_Model_Full}) as our model for non-Gaussianity,
i.e. we implement a uniform increase in the shot noise by a factor
$a$ by taking $n_{i}\to n_{i}/a$: 
\begin{equation}
C_{ab}^{\text{J}}\left(a\right)=C_{4,ab}^{\text{J}}+aC_{3,ab}^{\text{J}}+a^{2}C_{2,ab}^{\text{J}}\,.\label{eq:NG_Model_Jack}
\end{equation}
Although the Gaussian contributions $C_{4}^{\text{J}}$, $C_{3}^{\text{J}}$,
and $C_{2}^{\text{J}}$ are different, both qualitatively and quantitatively,
from their full-survey counterparts in in (\ref{eq:C4_Full}-\ref{eq:C2_Full}),
the same numerical methods that facilitate rapid integration of $C_{4}$,
$C_{3}$, and $C_{2}$ can be used on $C_{4}^{\text{J}}$, $C_{3}^{\text{J}}$,
and $C_{2}^{\text{J}}$. The restrictions required for the jackknife
model have been added to Rascal\footnote{\href{https://github.com/rcoconnell/Rascal}{https://github.com/rcoconnell/Rascal}}, the code used
in \cite{OConnell:2015src}. 

\begin{figure}
\includegraphics[width=2.1in]{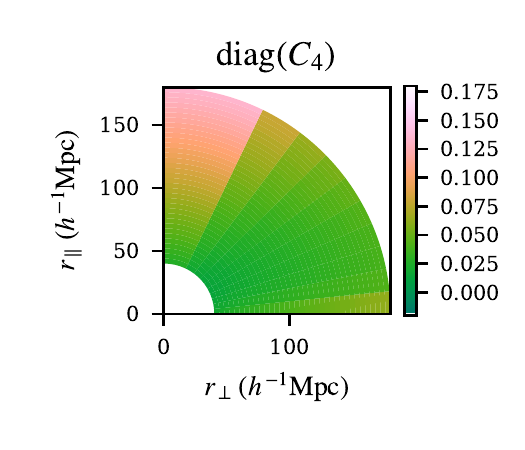}\includegraphics[width=2.1in]{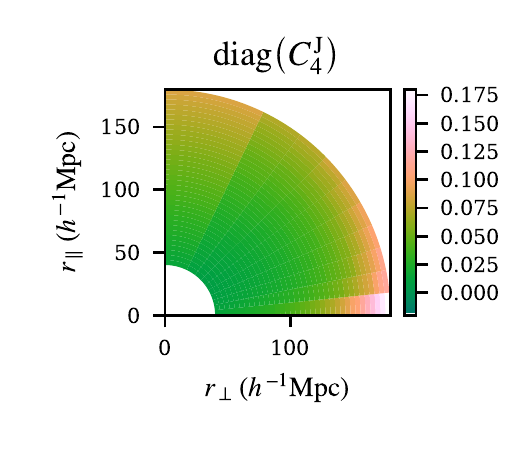}\includegraphics[width=2.1in]{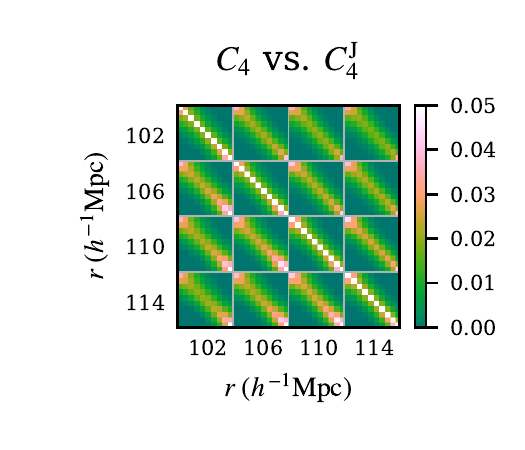}

\includegraphics[width=2.1in]{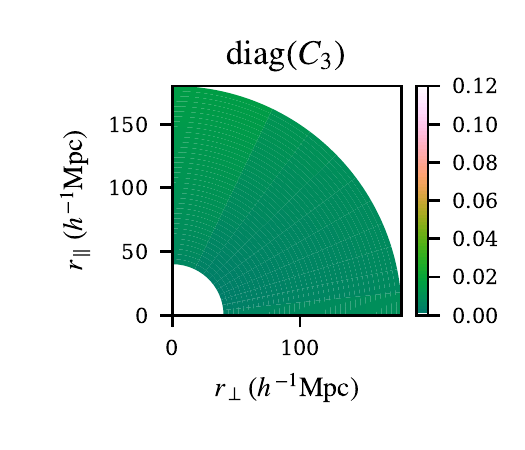}\includegraphics[width=2.1in]{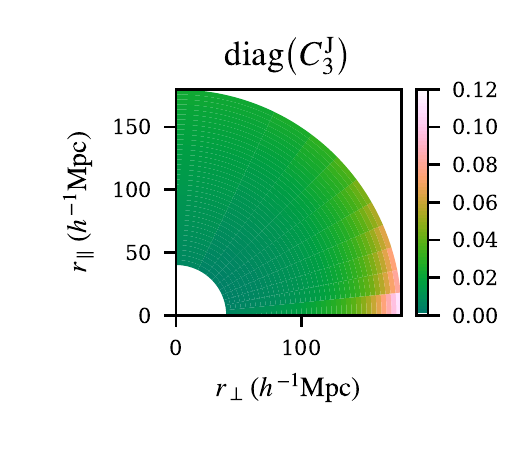}\includegraphics[width=2.1in]{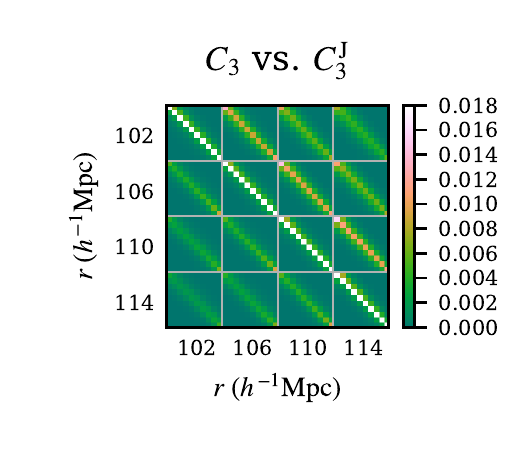}

\includegraphics[width=2.1in]{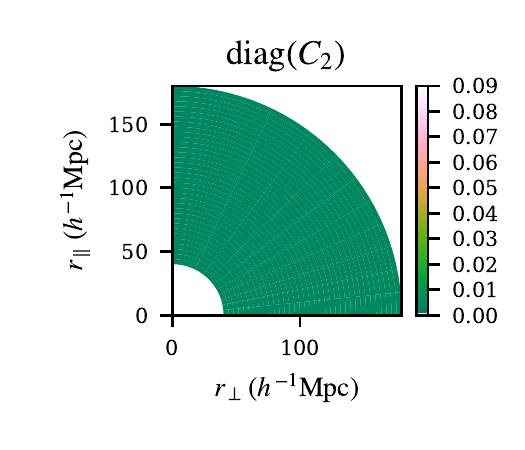}\includegraphics[width=2.1in]{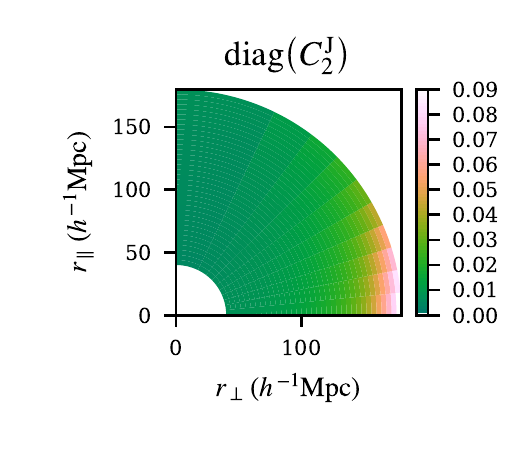}\hspace{2.1in}

\caption{\label{fig:Models}The components of the usual model covariance matrix
are $C_{4}$, $C_{3}$, and $C_{2}$, while for the jackknife covariance
they are $C_{4}^{\text{J}}$, $C_{3}^{\text{J}}$, and $C_{2}^{\text{J}}$.
In all plots the template is multiplied by $r_{a}r_{b}$ (in units
of $h^{-1}\text{Mpc}$) to remove the leading scaling in $r$. To
facilitate comparison we have separated out the diagonal components
and plotted them in terms of $r_{\parallel}$ and $r_{\perp}$. For
the off-diagonal elements we have plotted only a small portion of
the $350\times350$ matrix, with the usual model in the lower left
and the jackknife model in the upper right. In those plots we have
masked the diagonal entries, as they are faithfully represented in
the other plots. Note that $C_{2}$ and $C_{2}^{\text{J}}$ are proportional
to $\delta_{ab}$, i.e. they are diagonal.}

\end{figure}

To illustrate the differences between the Gaussian contributions to
the model with and without a jackknife, we have plotted each contribution
in Figure \ref{fig:Models}. The jackknife geometry is described in
detail in section \ref{subsec:validation-Jackknife}, but here we
mention that the jackknife regions are defined in angular coordinates
and that each jackknife region covers the entire redshift range of
the survey. Because the jackknife regions have limited extent in directions
transverse to the line-of-sight, the Gaussian contributions with jackknife
$C_{4}^{\text{J}}$, $C_{3}^{\text{J}}$, and $C_{2}^{\text{J}}$
are dramatically different from the Gaussian contributions without
jackknife $C_{4}$, $C_{3}$, and $C_{2}$. %

\section{Validation with Mocks}
\label{sec:Validation-Methods}

The approach we have described assumes that model covariance matrix
generation is performed in two steps: 
\begin{enumerate}
\item Generation of a family of models $C\left(a\right)$, using the survey
geometry, with a small number of unknown parameters (here one).
\item Estimation of those unknown parameter(s), here $a$. 
\end{enumerate}
In previous work \citep{OConnell:2015src} a sample covariance matrix,
determined from a large number of mock catalogues, was used to estimate
$a$. In this paper we propose to estimate $a$ using a jackknife
covariance matrix, as described in the previous section. We will consider
this new approach valid if it leads to estimates of $a$ that are
consistent with those that would arise from a sample analysis of a
large number of mocks. 

To perform this validation we use 1,000 mock catalogues. 900 of these
mocks are used to compute a sample covariance and establish a fiducial
value for $a$. For each of the 100 remaining mocks we will compute
a jackknife covariance matrix, then use that to produce an independent
estimate of $a$. We will then demonstrate that the 100 jackknife
estimates of $a$ are indeed consistent with the fiducial value.

We emphasise that this consistency is a non-trivial result. As shown
in Figure \ref{fig:RR}, the restricted jackknife covariance that
we use ``sees'' less volume, particularly for bins at large transverse
separation, than the sample covariance. If the preferred value of
the shot-noise rescaling $a$ (or other model parameters) were separation-dependent,
the resulting difference in weighting between small and large scales
for the jackknife versus the sample covariance could lead to inconsistent
estimates of $a$. Moreover, we saw in Figure \ref{fig:Models} that
the jackknife model, $C^{\text{J}}\left(a\right)$, is qualitatively
different from the full-volume model, $C\left(a\right)$, so without
some underlying physical understanding of $a$ we might not expect
fitting in the two approaches to lead to compatible estimates of $a$.
\begin{figure}
\includegraphics[width=2.1in]{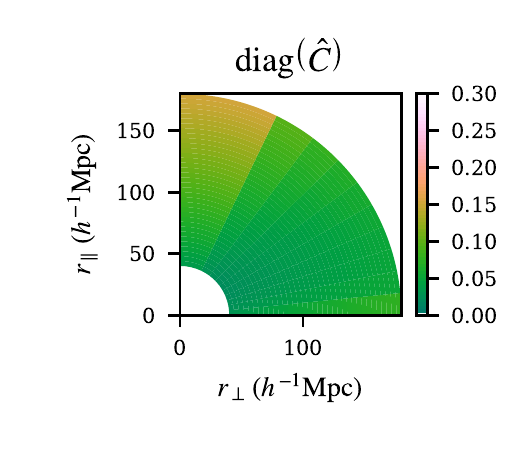}\includegraphics[width=2.1in]{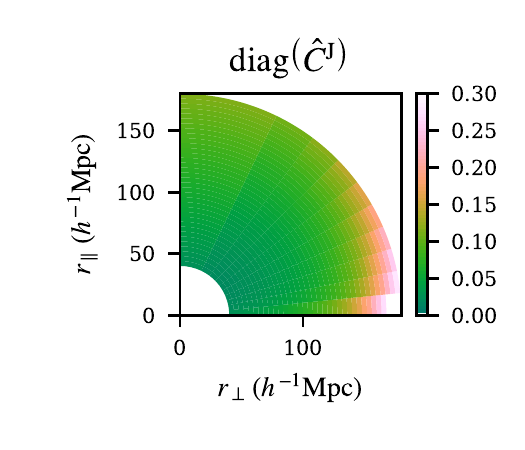}\includegraphics[width=2.1in]{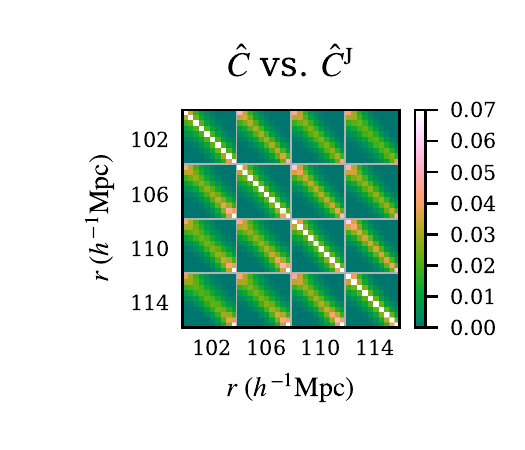}

\caption{\label{fig:Sample}Comparison of the sample covariance $\hat{C}$
computed from 900 QPM mocks and the jackknife covariance $\hat{C}^{\text{J}}$
computed from 100 QPM mocks. In all plots the covariance matrix is
multiplied by $r_{a}r_{b}$ (in units of $h^{-1}\text{Mpc}$), to
remove the leading scaling in $r$. To facilitate comparison we have
separated out the diagonal components (variances) and plotted them
in terms of $r_{\parallel}$ and $r_{\perp}$. For the off-diagonal
elements we have plotted only a small portion of the $350\times350$
matrix, with the sample covariance $\hat{C}$ in the lower left and
the jackknife covariance $\hat{C}^{\text{J}}$ in the upper right.
In that plot we have masked the diagonal entries, as they are faithfully
represented in the other plots. The jackknife covariance is qualitatively
different from the sample covariance, especially at large transverse
separations.}
\end{figure}

\subsection{Fitting Overview}

Following \cite{OConnell:2015src}, there are two likelihoods we
could use for covariance matrix fitting:
\begin{align}
-\log\mathcal{L}_{1}\left(a\right) & =\text{tr}\left[\Psi\left(a\right)\hat{C}\right]-\log\det\Psi\left(a\right).\\
-\log\mathcal{L}_{2}\left(a\right) & =\text{tr}\left[\hat{\Psi}C\left(a\right)\right]-\log\det C\left(a\right).
\end{align}
In these expressions $\hat{C}$ and $\hat{\Psi}$ are empirical covariance
and precision matrices, estimated from mocks, a jackknife, or by other
methods. $C\left(a\right)$ and $\Psi\left(a\right)$ are model covariance
and precision matrices, computed on the full volume or jackknife geometry,
as appropriate. Either likelihood can be minimised to provide an estimate
of $a$. While $\mathcal{L}_{1}$ and $\mathcal{L}_{2}$ are distinct
likelihoods, we expect them to yield compatible estimates of $a$
and the choice of which to use is largely a practical matter. The
primary practical consideration in choosing $\mathcal{L}_{1}$ or
$\mathcal{L}_{2}$ is our ability to accurately determine $\hat{\Psi}$
or $\Psi$$\left(a\right)$, given $\hat{C}$ or $C\left(a\right)$. 

When the empirical covariance matrix $\hat{C}$ is computed from independent
samples, as is the case with mock catalogues, the estimated matrix $\hat{C}$
follows a Wishart distribution and the inverse $\hat{\Psi}$ follows
an inverse-Wishart distribution. Wishart noise means that $\hat{C}^{-1}$
provides a biased estimate of the true precision matrix, but this
bias can be corrected. An unbiased estimate of $C^{-1}$ is provided
by:
\begin{align}
\hat{\Psi} & =\left(1-D\right)\hat{C}^{-1},\nonumber \\
D & =\frac{n_{\text{bins}}+1}{n_{\text{samples}}-1}.\label{eq:Wishart_Correction}
\end{align}
When $n_{\text{samples }}\gg n_{\text{bins}}$, as might be the case
when the samples in question are individual mock catalogues, it is straightforward
to compute $\hat{\Psi}$, and we might prefer to use $\mathcal{L}_{2}$
for fitting. Unfortunately an unbiased inverse is not available when
$n_{\text{samples}}\le n_{\text{bins}}+2$, and the matrix $\hat{C}$
is singular when $n_{\text{samples}}<n_{\text{bins}}$. This will
be the case in section \ref{subsec:validation-Jackknife}, where the
samples are jackknife regions and $\hat{C}$ is a jackknife covariance
matrix. In such cases we are forced to use $\mathcal{L}_{1}$, as
we will do for the remainder of this section.

In order to use $\mathcal{L}_{1}$ we must invert the model, $C\left(a\right)$.
This requires some care because $C\left(a\right)$ is
determined by numerical integration, and thus is noisy. In our case
the level of noise is very low, but because it is nonzero $C\left(a\right)^{-1}$
will provide a \emph{biased} estimate of $\Psi\left(a\right)$. Because
the noise on $C\left(a\right)$ is not Wishart-distributed, the simple
correction from (\ref{eq:Wishart_Correction}) does not apply. In
appendix \ref{sec:Inversion} we present a  partial solution to this
problem. The result, which we will use throughout this section, is
a corrected estimate of $\Psi\left(a\right)$ whose bias is shown
to be negligible for this application.

\subsection{Mock Catalogues and Jackknife}
\label{subsec:validation-Jackknife}

\begin{figure}
\begin{centering}
\includegraphics[width=3.25in]{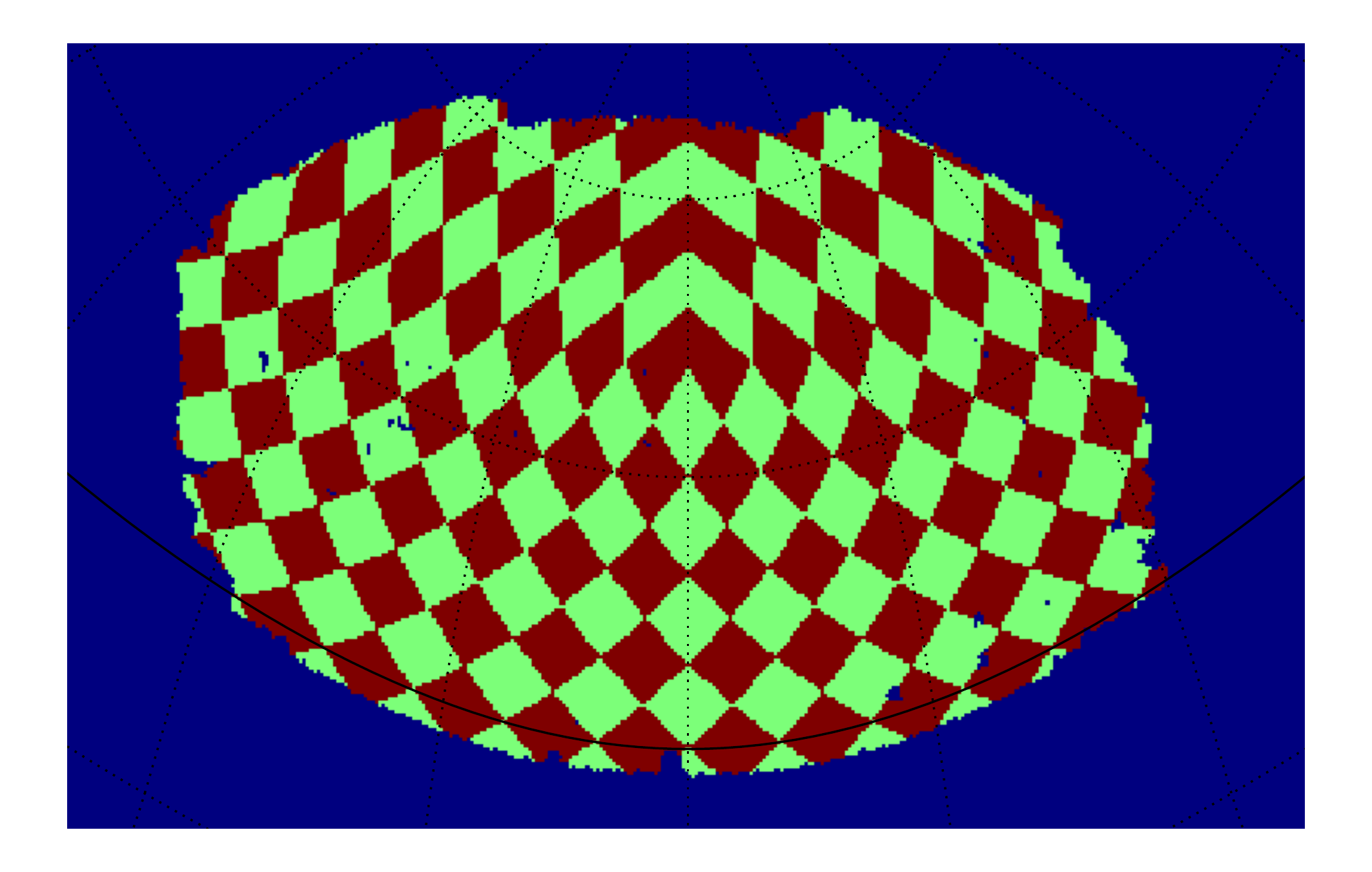}
\par\end{centering}
\caption{\label{fig:footprint}The survey footprint for the NGC portion of
the CMASS sample from BOSS. HealPix pixels with \texttt{nside=8},
used here as jackknife regions, are indicated with alternating colours.
While most regions are completely filled, some intersect the boundary
of the survey and are only partially filled.}
\end{figure}

To perform this validation we will use 1,000 quick particle mesh (QPM)
mocks \citep{White:2013psd} that match the NGC portion of the CMASS
sample from BOSS \citep{2013AJ....145...10D}. There are many schemes
that one could use to split this survey into jackknife regions. We
choose jackknife regions that are defined in terms of angular coordinates,
and cover the full redshift range of the sample. For the sake of simplicity
we use HEALPix pixels with \texttt{nside=8} \citep{2005ApJ...622..759G}.
When applied to the NGC portion of the CMASS sample of BOSS, this
divides the survey into 168 regions, with diameters at the midpoint
of the survey of $\sim180\,h^{-1}\text{Mpc}$. The survey footprint
and jackknife regions are shown in Figure \ref{fig:footprint}. 

\begin{figure}
\begin{centering}
\includegraphics[width=3.25in]{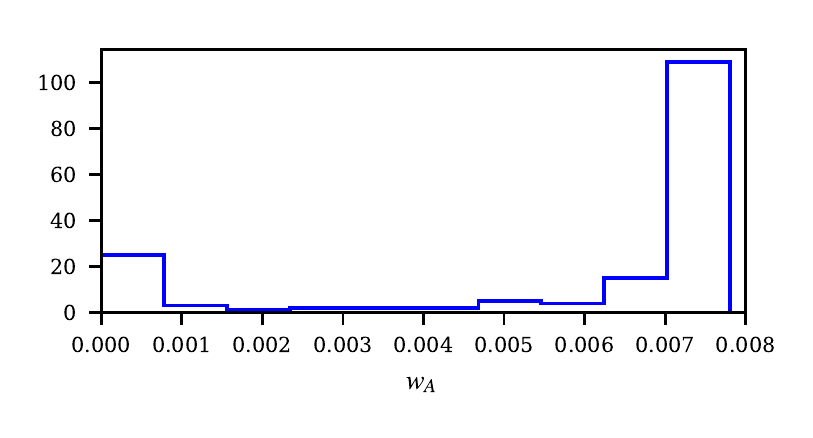}
\par\end{centering}
\caption{\label{fig:weights}A histogram of weights assigned to the 168 jackknife
regions. The bulk of regions receive the same weight, but a small
number that intersect the survey boundary receive less weight. This
heterogeneity is readily accommodated by a weighted jackknife.}
\end{figure}

By construction the pixels have equal areas, so jackknife regions
that are fully within the survey boundaries contain approximately
the same number of galaxies. Pixels that intersect the survey boundary
contain fewer galaxies, necessitating the weighting scheme introduced
in (\ref{eq:Region_Weights}). The distribution of weights for each
region (which is closely related to the region area) is shown in Figure
\ref{fig:weights}. 

We are using a 350 bin correlation function and so it is clear that
a jackknife covariance matrix for this correlation function, computed
using 168 jackknife regions, must be singular, as anticipated above. 

\subsection{Fitting Results}

As stated above, we use 900 of the QPM mocks to compute a sample covariance
matrix. We then fit this, using the $\mathcal{L}_{1}$ likelihood,
against our model $C\left(a\right)$ to make a single estimate of
the shot-noise rescaling parameter $a$. By leaving out individual
mocks we can make a jackknife estimate of the uncertainty in $a$,
with our final result $a=1.0590\pm0.0016$. This is the approach of
\cite{OConnell:2015src}, and the result will be used as the fiducial
value for $a$. 
\begin{table}
\begin{centering}
\begin{tabular}{|c|c|c|}
\hline 
Approach & $n_{\text{mocks}}$ & $a$\tabularnewline
\hline 
\hline 
Sample fitting & 900 & $1.0590\pm0.0016$\tabularnewline
\hline 
Jackknife fitting & 100 & $1.0597\pm0.0009$\tabularnewline
\hline 
\end{tabular}
\par\end{centering}
\caption{\label{tab:Fitting_Results}Fitting results from 1,000 QPM mocks,
including $1\sigma$ uncertainties. In the ``Sample fitting'' approach
900 mocks are used to compute a single sample covariance, which is
then used to fit a model precision matrix $\Psi\left(a\right)$. In
the ``Jackknife fitting'' approach 100 mocks are used to generate
100 independent jackknife covariance matrices, and a model precision
matrix $\Psi^{\text{J}}\left(a\right)$ is fit against each separately.
Fits against a \emph{single} jackknife covariance matrix would give
(on average) $a=1.0597\pm0.0086$.}
\end{table}

\begin{figure}
\begin{centering}
\includegraphics[width=3.25in]{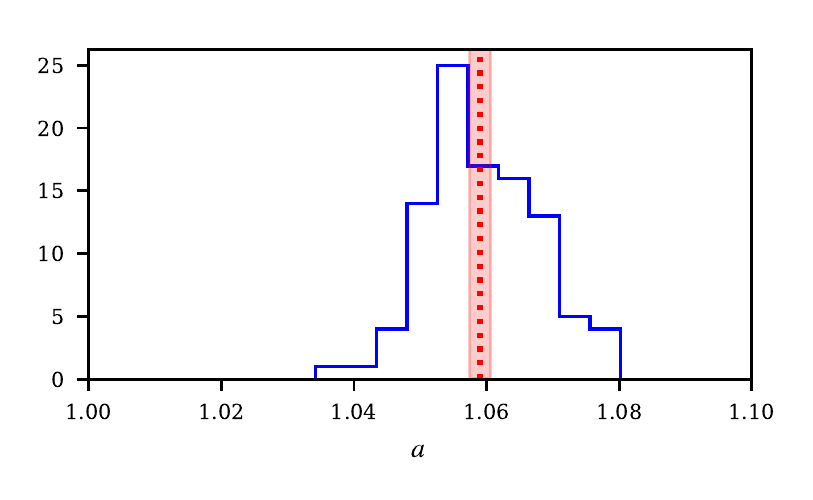}
\par\end{centering}
\caption{\label{fig:jackknife_a}The results of fitting the shot-noise rescaling
$a$ from 100 QPM mocks. For each individual mock we compute a jackknife
covariance, then fit that to a model for the jackknife covariance.
The results are in excellent agreement with the value of $a=1.0590\pm0.0016$
determined from a sample covariance computed using 900 QPM mocks,
indicated here by the dotted red line and shaded red band.}
\end{figure}

For each of the remaining 100 mocks we computed the jackknife covariance
matrix $\hat{C}_{ab}^{\text{J}}$ using (\ref{eq:Jackknife_C}), then
used the $\mathcal{L}_{1}$ likelihood to fit the jackknife model
$C^{\text{J}}\left(a\right)$ and estimate $a$. The result is 100
independent estimates of $a$. A histogram of these 100 estimates
is provided in Figure \ref{fig:jackknife_a}. Normal sample statistics
on those estimates give $a=1.0597\pm0.0009$. Note that 0.0009 is
the error on the mean, and a jackknife fit using a single mock would
give (on average) $a=1.0597\pm0.0086$. The fitting results are summarised
in \ref{tab:Fitting_Results}. We find that the two fitting methods
lead to compatible estimates of $a$, validating the jackknife approach.

One surprising aspect of these fitting results is that the jackknife
analysis of 100 mocks provides considerably more information about
the shot-noise rescaling $a$ than does a sample analysis of 900 mocks.
Intuitively, each jackknife region is large enough to provide information
on its own about $a$, and some of that information is lost when regions
are combined in the sample analysis. 

\begin{figure}
\includegraphics[width=2.1in]{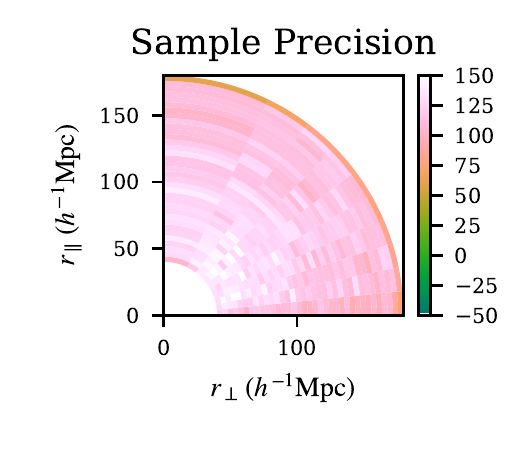}\includegraphics[width=2.1in]{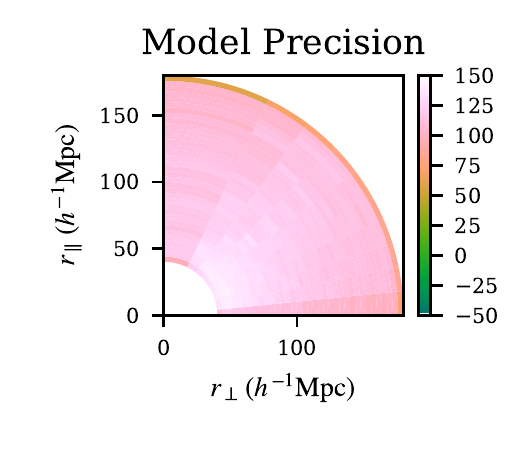}\includegraphics[width=2.1in]{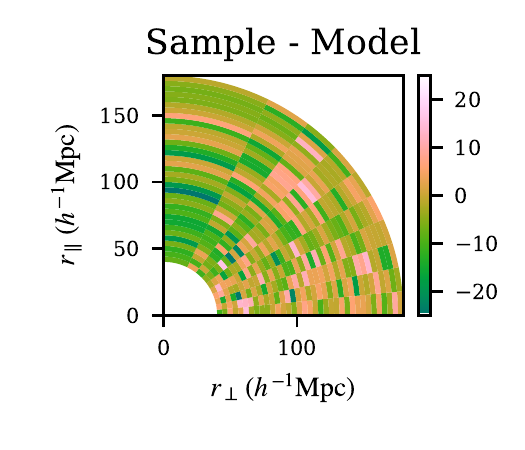}

\includegraphics[width=2.1in]{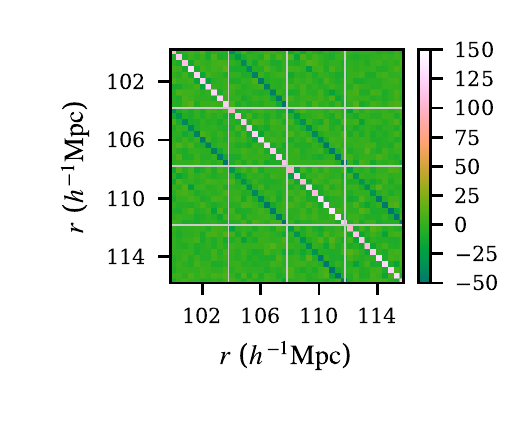}\includegraphics[width=2.1in]{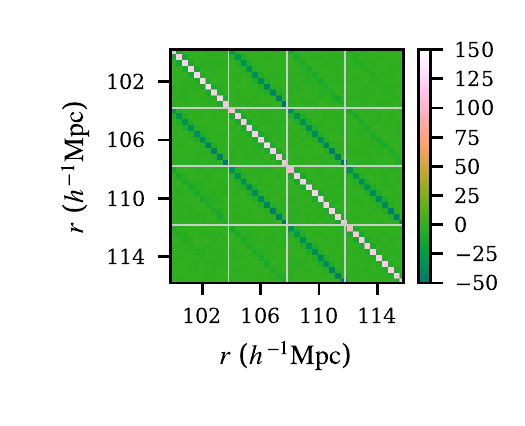}\includegraphics[width=2.1in]{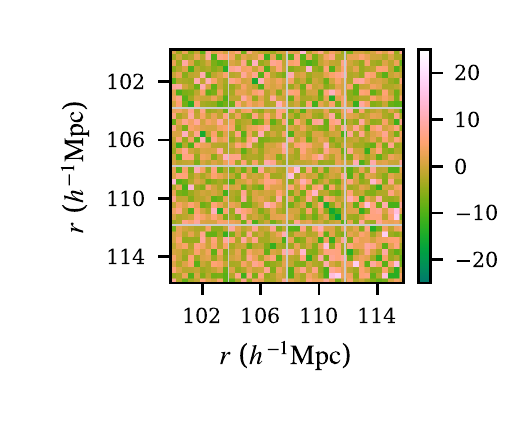}

\caption{\label{fig:Precision}Comparison of the sample precision matrix computed
from 900 QPM mocks\textbf{ }with the model precision matrix calibrated
using the other 100 QPM mocks. We focus on the precision matrix rather
than the covariance matrix because the precision matrix is used to
construct likelihoods. We plot $\hat{\Psi}_{ab}/r_{a}r_{b}$ and $\Psi\left(a\right)/r_{a}r_{b}$
to remove the leading scaling of the precision matrix, and only include
a small portion of the $350\times350$ bin matrix so that some detail
is visible. Our value of $a=1.0597$ is determined from fits against
jackknife covariance matrices, rather than direct calibration using
the sample covariance. The difference between the sample precision
and the model appears to be consistent with noise, indicating that
our jackknife calibration method has been successful. }

\end{figure}

In Figure \ref{fig:Precision} we present the sample precision matrix
$\hat{\Psi}$, computed from 900 QPM mocks, and the model precision
matrix $\Psi^{\text{NG}}\left(a\right)$, with $a=1.0597$ as was
found from the \emph{jackknife} fits of the other 100 QPM mocks. The
difference between the two matrices appears to be consistent with
noise, providing additional evidence that the jackknife method has
been successful. 

\subsection{Error Estimation for the Jackknife}
\label{subsec:Validation-Error}

We have demonstrated that the jackknife method can provide an accurate
estimate of the shot-noise rescaling $a$ from a single survey volume.
In situations where we have done this, it might be interesting to
estimate the uncertainty in the estimate of $a$. An additional jackknife
can provide a reliable estimate of the uncertainty in $a$, even with
a single survey volume.

Recall that $w_{aA}$ is the weight assigned to jackknife region $A$
when considering correlation function bin $a$ and that $\hat{\xi}_{aA}$
is the estimate of the correlation function in bin $a$ from region
$A$ alone. We can estimate the covariance matrix with a single region
$B$ excluded as follows. First, we recompute our weights and the
average correlation function,
\begin{align}
w_{aA,B} & =\frac{w_{aA}}{\sum_{C\neq B}w_{aC}}\,,\\
\hat{\xi}_{a,B}^{\text{J}} & =\sum_{A\neq B}w_{aA,B}\hat{\xi}_{aA}\,.
\end{align}
then use the updated weights to compute the weighted jackknife covariance, with the single region $B$ omitted,
\begin{equation}
\hat{C}_{ab,B}^{\text{J}}=\frac{1}{1-\sum_{C\neq B}w_{aC,B}w_{bC,B}}\left[\sum_{A\neq B}w_{aA,B}w_{bA,B}\left(\hat{\xi}_{aA,B}-\hat{\xi}_{a}^{\text{J}}\right)\left(\hat{\xi}_{bA,B}-\hat{\xi}_{b}^{\text{J}}\right)\right].
\end{equation}
As in \eqref{eq:Jackknife_C} the usual jackknife prescription in terms of dropped regions reduces to a rescaled sample covariance.
We then fit the model against $\hat{C}_{ab,B}^{\text{J}}$ using the
$\mathcal{L}_{1}$ likelihood to find an estimate of the shot-noise
rescaling, $a_{B}$, with region $B$ left out. 

Our next goal is to combine the $a_{B}$ into a jackknife estimate
of $\text{var}\left(a\right)$ obtained from the single survey volume.
This first requires that we convert our bin-dependent weights, $w_{aA}$,
into weights associated with the jackknife region alone, $w_{A}$.
A simple mean computed across the bins used in the analysis accomplishes
this. We then perform a weighted jackknife to estimate $\sigma_{a}^{\text{J}}$:
\begin{align}
\overline{a} & =\sum_{A}w_{A}a_{A}\,,\\
\sigma_{a}^{\text{J}} & =\sqrt{\frac{1}{1-\sum_{B}w_{B}^{2}}\sum_{A}\left(1-w_{A}\right)^{2}\left(a_{A}-\overline{a}\right)^{2}}.
\end{align}
As the jackknife regions are not independent, we expect $\sigma_{a}^{J}$
to provide a biased estimate of the true uncertainty $\sigma_{a}$.
When we perform this analysis on 100 QPM mocks and average the results,
we find 
\begin{equation}
\sigma_{a}^{\text{J}}=0.0082\pm0.0006\,.
\end{equation}
If we instead compute a sample variance on the values of $a$ computed
from jackknife fits for those same 100 mocks we find $\sigma_{a}=0.0086$,
in surprisingly good agreement with the jackknife error estimate $\sigma_{a}^{\text{J}}$.
This suggests that the jackknife error estimate provides a useful
way to characterise the error on $a$ when it is estimated from a
single survey volume.

\subsection{Consequences for Measurement}

So far we have focused on our ability to accurately and precisely
estimate the shot-noise rescaling parameter $a$. We now focus on
the implications of that estimate for a hypothetical measurement of
cosmological parameters. The primary complication is that measurements
will involve a variety of modes \textendash{} not just those associated
with the cosmological parameters themselves, but also modes associated
with additional nuisance parameters. Rather than choose a particular
set of modes to examine, we focus on the average variance contributed
by a single mode, 
\begin{equation}
\gamma\left(C\right)=\left[\text{det}\left(C\right)\right]^{1/n_{\text{bins}}}.
\end{equation}
Under a uniform rescaling, 
\begin{equation}
\gamma\left(\alpha C\right)=\alpha\gamma\left(C\right).
\end{equation}
While the changes to the covariance matrix that we will consider are
\emph{not} uniform, we generally expect that when $\gamma$ increases
 the parameter covariance matrix will increase as well, with changes in $\gamma$ tied to changes in the parameter covariance matrix at the order-of-magnitude level.

We first consider whether the value of $a$ that we have determined,
$a=1.0597$, is sufficiently different from $a=1$ to be of interest.
If we look at the average variance, we find 
\begin{align}
\gamma\left(C\left(a=1.0597\right)\right) & =1.087\times10^{-6}\,,\\
\gamma\left(C\left(a=1\right)\right) & =0.998\times10^{-6}\,.
\end{align}
That is, if we ignore the shot-noise rescaling while performing a
measurement, we expect the resulting parameter covariance matrix to
be too small by roughly 9\%. This establishes the necessity of applying
the shot-noise rescaling.

The uncertainty in our estimate of $a$ will propagate through to uncertainty in the the covariance matrix. Using $\gamma$ to measure this, we find that an uncertainty in $a$ of $\sigma_a = 0.009$, as we would expect from a jackknife estimate using a single survey volume, corresponds to an uncertainty in $\gamma$ of $\sigma_\gamma = 1.4\times 10^{-8}$. In other words, for the BOSS-like survey considered here a 15\% measurement of $a-1$ contributes roughly 1\%  to the error-on-the-error. At this point a variety of competing concerns about the error-on-the-error arise, including our approximation of contributions from the connected three- and four-point functions by additional shot noise, so we find the single-volume measurement of $a$ to be sufficiently accurate.

\section{Outlook}
\label{sec:Outlook}

Here we have generated a model covariance matrix for a BOSS-like survey.
By construction, that model accurately reflects the long-distance physics and geometry
of the survey. Short-distance physics is modelled
by a single shot-noise rescaling parameter, and we have demonstrated
that this parameter can be accurately and precisely calibrated using
a single survey volume. The jackknife methods we have used for this calibration do not require significant computational resources beyond those required to pair-count the single survey volume. While we have validated this procedure using
mock catalogues, we anticipate that future applications will use the
actual survey as the single survey volume, obviating the need to use
mock catalogues for covariance matrix estimation.

We emphasise that mock catalogues still play a vital role in estimating
the size of systematic uncertainties. When generating mock catalogues there are clear trade-offs between accuracy and quantity. We
hope that this method will allow limited computing resources to be
focused on the former, by alleviating the need for the latter.

The larger lesson from our study is that because the shot-noise rescaling
$a$ is determined by short-distance physics, it need not be estimated
using covariance matrices that reflect the full survey geometry. While
we anticipate that the jackknife approach will be very convenient,
one can also imagine applications where several small-volume cubic
mocks are used to estimate $a$. We can even imagine applying the
jackknife approach to those small volumes in order to increase the
precision with which $a$ can be estimated.

While we have focused on a simple model for the short-distance physics
of the survey, we do not expect our results to apply to this model
alone. More accurate models of the short-distance physics might include
several parameters governing redshift-space distortions and the connected
3- and 4-point functions. We expect the fitting methods described
here could be readily applied to these models, with the caveat that
parameters that are physically distinct could have degenerate, or
nearly degenerate, impacts on the covariance matrix. 
The opportunity to fit such models either to the survey data themselves or to small-volume cubic mock catalogues could substantially reduce the computational requirements for next-generation surveys by removing the need to generate and analyse thousands of full-volume mock catalogues.

\section*{Acknowledgements}
DJE is supported by U.S. Department of Energy grant DE-SC0013718 and as a Simons Foundation investigator.

\bibliographystyle{mnras}
\bibliography{CovRefs}

\begin{thebibliography}{}
\makeatletter
\relax
\def\mn@urlcharsother{\let\do\@makeother \do\$\do\&\do\#\do\^\do\_\do\%\do\~}
\def\mn@doi{\begingroup\mn@urlcharsother \@ifnextchar [ {\mn@doi@}
  {\mn@doi@[]}}
\def\mn@doi@[#1]#2{\def\@tempa{#1}\ifx\@tempa\@empty \href
  {http://dx.doi.org/#2} {doi:#2}\else \href {http://dx.doi.org/#2} {#1}\fi
  \endgroup}
\def\mn@eprint#1#2{\mn@eprint@#1:#2::\@nil}
\def\mn@eprint@arXiv#1{\href {http://arxiv.org/abs/#1} {{\tt arXiv:#1}}}
\def\mn@eprint@dblp#1{\href {http://dblp.uni-trier.de/rec/bibtex/#1.xml}
  {dblp:#1}}
\def\mn@eprint@#1:#2:#3:#4\@nil{\def\@tempa {#1}\def\@tempb {#2}\def\@tempc
  {#3}\ifx \@tempc \@empty \let \@tempc \@tempb \let \@tempb \@tempa \fi \ifx
  \@tempb \@empty \def\@tempb {arXiv}\fi \@ifundefined
  {mn@eprint@\@tempb}{\@tempb:\@tempc}{\expandafter \expandafter \csname
  mn@eprint@\@tempb\endcsname \expandafter{\@tempc}}}

\bibitem[\protect\citeauthoryear{{Bernstein}}{{Bernstein}}{1994}]{Bernstein1994}
{Bernstein} G.~M.,  1994, \mn@doi [Astrophysical Journal] {10.1086/173915},
  \href {http://adsabs.harvard.edu/abs/1994ApJ...424..569B} {424, 569}

\bibitem[\protect\citeauthoryear{Chuang et~al.}{Chuang
  et~al.}{2015}]{Chuang:2014toa}
Chuang C.-H.,  et~al., 2015, \mn@doi [Mon. Not. Roy. Astron. Soc.]
  {10.1093/mnras/stv1289}, 452, 686

\bibitem[\protect\citeauthoryear{{Dawson} et~al.,}{{Dawson}
  et~al.}{2013}]{2013AJ....145...10D}
{Dawson} K.~S.,  et~al., 2013, \mn@doi [Astrophysical Journal]
  {10.1088/0004-6256/145/1/10}, \href
  {http://adsabs.harvard.edu/abs/2013AJ....145...10D} {145, 10}

\bibitem[\protect\citeauthoryear{Dodelson \& Schneider}{Dodelson \&
  Schneider}{2013}]{Dodelson:2013uaa}
Dodelson S.,  Schneider M.~D.,  2013, \mn@doi [Phys.Rev.]
  {10.1103/PhysRevD.88.063537}, D88, 063537

\bibitem[\protect\citeauthoryear{Dor{\'e} et~al.}{Dor{\'e}
  et~al.}{2018}]{Dore:2018smn}
Dor{\'e} O.,  et~al., 2018, preprint (\mn@eprint {arXiv} {1804.03628})

\bibitem[\protect\citeauthoryear{Escoffier et~al.,}{Escoffier
  et~al.}{2016}]{Escoffier:2016qnf}
Escoffier S.,  et~al., 2016, preprint (\mn@eprint {arXiv} {1606.00233})

\bibitem[\protect\citeauthoryear{{G{\'o}rski}, {Hivon}, {Banday}, {Wandelt},
  {Hansen}, {Reinecke}  \& {Bartelmann}}{{G{\'o}rski}
  et~al.}{2005}]{2005ApJ...622..759G}
{G{\'o}rski} K.~M.,  {Hivon} E.,  {Banday} A.~J.,  {Wandelt} B.~D.,  {Hansen}
  F.~K.,  {Reinecke} M.,   {Bartelmann} M.,  2005, \mn@doi [Astrophysical
  Journal] {10.1086/427976}, \href
  {http://adsabs.harvard.edu/abs/2005ApJ...622..759G} {622, 759}

\bibitem[\protect\citeauthoryear{Grieb, S{\'a}nchez, Salazar-Albornoz  \&
  Dalla~Vecchia}{Grieb et~al.}{2016}]{Grieb:2015bia}
Grieb J.~N.,  S{\'a}nchez A.~G.,  Salazar-Albornoz S.,   Dalla~Vecchia C.,
  2016, \mn@doi [Mon. Not. Roy. Astron. Soc.] {10.1093/mnras/stw065}, 457, 1577

\bibitem[\protect\citeauthoryear{Howlett \& Percival}{Howlett \&
  Percival}{2017}]{Howlett:2017vwp}
Howlett C.,  Percival W.~J.,  2017, \mn@doi [Mon. Not. Roy. Astron. Soc.]
  {10.1093/mnras/stx2342}, 472, 4935

\bibitem[\protect\citeauthoryear{Joachimi}{Joachimi}{2017}]{Joachimi:2016xhk}
Joachimi B.,  2017, \mn@doi [Mon. Not. Roy. Astron. Soc.]
  {10.1093/mnrasl/slw240}, 466, L83

\bibitem[\protect\citeauthoryear{Klypin \& Prada}{Klypin \&
  Prada}{2018}]{Klypin:2017}
Klypin A.,  Prada F.,  2018, \mn@doi [Monthly Notices of the Royal Astronomical
  Society] {10.1093/mnras/sty1340}, 478, 4602

\bibitem[\protect\citeauthoryear{{Laureijs} et~al.,}{{Laureijs}
  et~al.}{2011}]{EuclidRedBook}
{Laureijs} R.,  et~al., 2011, preprint, \href
  {http://adsabs.harvard.edu/abs/2011arXiv1110.3193L} {} (\mn@eprint {arXiv}
  {1110.3193})

\bibitem[\protect\citeauthoryear{Levi et~al.}{Levi et~al.}{2013}]{Levi:2013gra}
Levi M.,  et~al., 2013, preprint (\mn@eprint {arXiv} {1308.0847})

\bibitem[\protect\citeauthoryear{Lippich et~al.}{Lippich
  et~al.}{2018}]{Lippich:2018wrx}
Lippich M.,  et~al., 2018, preprint (\mn@eprint {arXiv} {1806.09477})

\bibitem[\protect\citeauthoryear{O'Connell, Eisenstein, Vargas, Ho  \&
  Padmanabhan}{O'Connell et~al.}{2016}]{OConnell:2015src}
O'Connell R.,  Eisenstein D.,  Vargas M.,  Ho S.,   Padmanabhan N.,  2016,
  \mn@doi [Mon. Not. Roy. Astron. Soc.] {10.1093/mnras/stw1821}, 462, 2681

\bibitem[\protect\citeauthoryear{Padmanabhan, White, Zhou  \&
  O'Connell}{Padmanabhan et~al.}{2016}]{Padmanabhan:2015vlf}
Padmanabhan N.,  White M.,  Zhou H.~H.,   O'Connell R.,  2016, \mn@doi [Mon.
  Not. Roy. Astron. Soc.] {10.1093/mnras/stw1042}, 460, 1567

\bibitem[\protect\citeauthoryear{Pearson \& Samushia}{Pearson \&
  Samushia}{2016}]{Pearson:2015gca}
Pearson D.~W.,  Samushia L.,  2016, \mn@doi [Mon. Not. Roy. Astron. Soc.]
  {10.1093/mnras/stw062}, 457, 993

\bibitem[\protect\citeauthoryear{Percival et~al.}{Percival
  et~al.}{2014}]{Percival:2013sga}
Percival W.~J.,  et~al., 2014, \mn@doi [Monthly Notices of the Royal
  Astronomical Society] {10.1093/mnras/stu112}, 439, 2531

\bibitem[\protect\citeauthoryear{Vargas-Maga{\~n}a et~al.,}{Vargas-Maga{\~n}a
  et~al.}{2018}]{Vargas-2016}
Vargas-Maga{\~n}a M.,  et~al., 2018, \mn@doi [Monthly Notices of the Royal
  Astronomical Society] {10.1093/mnras/sty571}, 477, 1153

\bibitem[\protect\citeauthoryear{White, Tinker  \& McBride}{White
  et~al.}{2014}]{White:2013psd}
White M.,  Tinker J.~L.,   McBride C.~K.,  2014, \mn@doi [Monthly Notices of
  the Royal Astronomical Society] {10.1093/mnras/stt2071}, 437, 2594

\bibitem[\protect\citeauthoryear{Wishart}{Wishart}{1928}]{Wishart1928}
Wishart J.,  1928, Biometrika, 20A, 32

\bibitem[\protect\citeauthoryear{Zhu, Padmanabhan  \& White}{Zhu
  et~al.}{2015}]{Zhu:2014ica}
Zhu F.,  Padmanabhan N.,   White M.,  2015, \mn@doi [Monthly Notices of the
  Royal Astronomical Society] {10.1093/mnras/stv964}, 451, 4755

\bibitem[\protect\citeauthoryear{Zhu et~al.}{Zhu et~al.}{2018}]{Zhu:2018edv}
Zhu F.,  et~al., 2018, preprint (\mn@eprint {arXiv} {1801.03038})

\makeatother
\end{thebibliography}

\appendix

\section{Reducing Bias in Estimates of Precision Matrices}
\label{sec:Inversion}

Although we often speak about different methods to estimate the covariance
matrix, data analysis typically requires that we invert that covariance
in order to estimate the precision matrix. Quite generally, noise
on the estimate of the covariance matrix leads to bias in the estimate
of the precision matrix. Suppose that $\hat{C}$ provides an estimate
of a covariance matrix $C$, with some noise $N$:
\begin{equation}
\hat{C}=C+N\,.
\end{equation}
We depart slightly from the notation in the body of the paper and
have in mind that the noisy estimate $\hat{C}$ could be a sample
covariance, jackknife estimate, \emph{or} a noisy model. Our goal
is to estimate $C^{-1}$, with the assumption that $\left\langle N\right\rangle =0$. 

If we invert $\hat{C}$ we find 
\begin{equation}
\hat{C}^{-1}=C^{-1}-C^{-1}NC^{-1}+\left(C^{-1}N\right)^{2}C^{-1}+\mathcal{O}\left(N^{3}\right).
\end{equation}
The assumption that $\left\langle N\right\rangle =0$, eliminates
the $\mathcal{O}\left(N\right)$ term, but the term at $\mathcal{O}\left(N^{2}\right)$
is nonzero in expectation:
\begin{equation}
\left\langle \hat{C}^{-1}\right\rangle =C^{-1}+\left\langle \left(C^{-1}N\right)^{2}\right\rangle C^{-1}+\dots\label{eq:C_inverse_series}
\end{equation}
When $\hat{C}$ is a sample covariance matrix it follows the well-understood
Wishart distribution \citep{Wishart1928}, and the series in (\ref{eq:C_inverse_series})
can be resummed to give 
\begin{align}
\left\langle \hat{C}^{-1}\right\rangle  & =\left(1-D\right)^{-1}C^{-1}\,,\label{eq:Wishart_Again}\\
D & =\frac{n_{\text{bins}}+1}{n_{\text{samples}}-1}\,.
\end{align}
It follows that $\left(1-D\right)\hat{C}^{-1}$ provides an unbiased
estimate of $C^{-1}$. When $\hat{C}$ is \emph{not} a sample covariance
we do not expect it to follow the Wishart distribution, and require
some other approach to manage the bias on $\left\langle \hat{C}^{-1}\right\rangle $.
In the following we introduce a jackknife-inspired approach which
can reduce (but not eliminate) this bias.

\subsection{Quadratic Correction}

We begin by assuming that rather than a single estimate of the covariance
matrix, there are $n$ independent estimates:
\begin{equation}
\hat{C}_{i}=C+N_{i}\,.
\end{equation}
In this paper the estimate is a result of numerical integration, so
this is simply a matter of dividing the time dedicated to numerical integration among several independent runs, rather than using it for a single run. The most precise estimate of the model is then 
\[
\hat{C}=\frac{1}{n}\sum_{i=1}^{n}\hat{C}_{i}\,.
\]
As above, $\left\langle \hat{C}^{-1}\right\rangle $ will be biased,
but the multiple independent estimates will allow us to perform a
separate estimate of this bias and subtract it off.

The inverse of $\hat{C}$ is 
\begin{align}
\hat{C}^{-1} & =C^{-1}-C^{-1}\left(\frac{1}{n}\sum_{i=1}^{n}N_{i}\right)C^{-1}+\left[C^{-1}\left(\frac{1}{n}\sum_{i=1}^{n}N_{i}\right)\right]^{2}C^{-1}+\dots\\
 & =C^{-1}-C^{-1}\left(\frac{1}{n}\sum_{i=1}^{n}N_{i}\right)C^{-1}+\frac{1}{n^{2}}\sum_{i}\left(C^{-1}N_{i}\right)^{2}C^{-1}\nonumber \\
 & \hphantom{=C^{-1}}+\frac{1}{n^{2}}\sum_{i\neq j}\left(C^{-1}N_{i}\right)\left(C^{-1}N_{j}\right)C^{-1}
\end{align}
If the estimates are unbiased and independent, we have 
\begin{align}
\left\langle N_{i}\right\rangle  & =0\,,\\
\left\langle N_{i}N_{j}\right\rangle  & =0,\,i\neq j\,,
\end{align}
and so 
\begin{equation}
\left\langle \hat{C}^{-1}\right\rangle =C^{-1}+\frac{1}{n^{2}}C^{-1}\left\langle \sum_{i}N_{i}C^{-1}N_{i}\right\rangle C^{-1}+\dots\label{eq:C_inverse_bias}
\end{equation}
We wish to estimate the quadratic bias. To do this we introduce 
\begin{align}
\hat{C}_{\left[i\right]} & =\frac{1}{n-1}\sum_{j\neq i}\hat{C}_{j}\\
 & =C+\frac{1}{n-1}\sum_{j\neq i}N_{j}\,.
\end{align}
Its inverse is 
\begin{equation}
\hat{C}_{\left[i\right]}^{-1}=C^{-1}-C^{-1}\left(\frac{1}{n-1}\sum_{j\neq i}N_{j}\right)C^{-1}+\left[C^{-1}\left(\frac{1}{n-1}\sum_{j\neq i}N_{j}\right)\right]^{2}C^{-1}+\dots\,.
\end{equation}
Our ultimate goal is to isolate the quadratic term. Toward that end
it is beneficial to have a series that starts at $1$, rather than
$C^{-1}$, so we multiply by $\hat{C_{i}}$:
\begin{align}
\hat{C}_{\left[i\right]}^{-1}\hat{C}_{i} & =1-C^{-1}\left(\frac{1}{n-1}\sum_{j\neq i}N_{j}\right)+C^{-1}N_{i}+\left[C^{-1}\left(\frac{1}{n-1}\sum_{j\neq i}N_{j}\right)\right]^{2}\nonumber \\
 & \hphantom{=1}-C^{-1}\left(\frac{1}{n-1}\sum_{j\neq i}N_{j}\right)C^{-1}N_{i}+\dots
\end{align}
If we then average over $i$, we find that the linear terms cancel:
\begin{equation}
\frac{1}{n}\sum_{i}\hat{C}_{\left[i\right]}^{-1}\hat{C}_{i}=1+\frac{1}{n\left(n-1\right)}\sum_{i}\left(C^{-1}N_{i}\right)^{2}-\frac{1}{n\left(n-1\right)^{2}}\sum_{i\neq j}\left(C^{-1}N_{i}\right)\left(C^{-1}N_{j}\right)+\dots
\end{equation}
In expectation, we have 
\begin{equation}
\frac{1}{n}\left\langle \sum_{i}\hat{C}_{\left[i\right]}^{-1}\hat{C}_{i}\right\rangle =1+\frac{1}{n\left(n-1\right)}\sum_{i}C^{-1}\left\langle N_{j}C^{-1}N_{j}\right\rangle +\dots
\end{equation}
and we recognise the same quadratic contribution as appeared in (\ref{eq:C_inverse_bias}).

To actually perform the correction we introduce
\begin{equation}
\widetilde{D}=\frac{n-1}{n}\left[-1+\frac{1}{n}\sum_{i}\hat{C}_{\left[i\right]}^{-1}\hat{C}_{i}\right]\,.
\end{equation}
The coefficients are chosen so that 
\begin{equation}
\left\langle \left(1-\widetilde{D}\right)\hat{C}^{-1}\right\rangle =C^{-1}+\mathcal{O}\left(N^{3}\right),
\end{equation}
so the quadratic contribution to the bias is eliminated. Note that
the only assumptions we have made are that $\left\langle N\right\rangle =0$
and that the estimates $\hat{C}_{i}$ are independent from one another.
Since the method does not rest on assumptions about the distribution
of the noise, we expect it to be applicable to a wide variety of non-sample
covariance matrices, not just the model described in this paper.

\subsection{Application to our Model Precision Matrix}

To illustrate this method of inversion we apply it to the model computed
in section (\ref{subsec:Review-Non-Gaussianity}) of the main body
of the paper. That approximates the covariance matrix for the 2-point
galaxy correlation function in a BOSS-like survey, with the correlation
function estimated in 350 bins of $\Delta r=4\,h^{-1}\text{Mpc}$
and $\Delta\mu=0.1$. We generated $n=10$ independent estimates of
the model covariance matrix.

\begin{figure}
\includegraphics{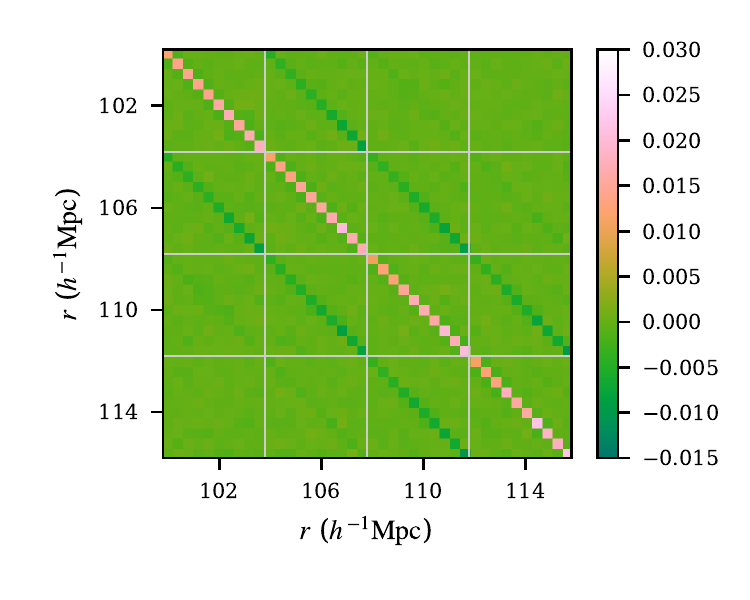}\hfill{}\includegraphics{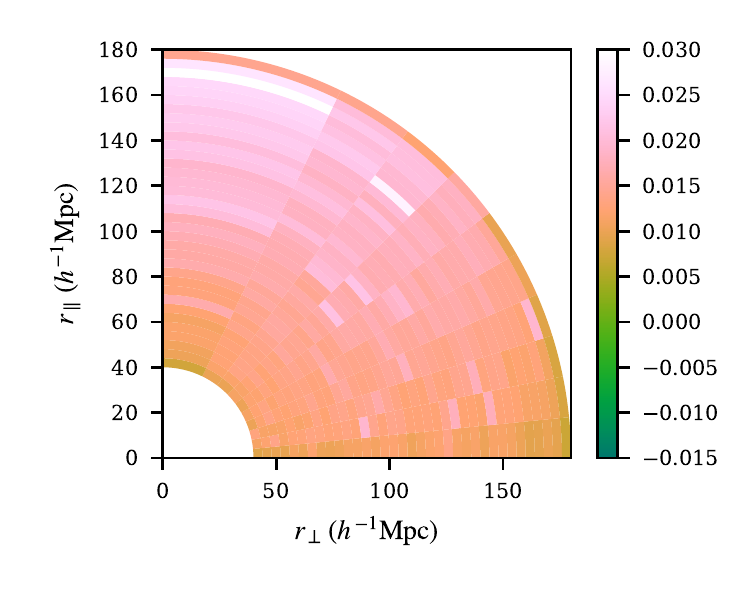}

\caption{\label{fig:D}On the left we plot a small portion of the $\widetilde{D}$
matrix. We observe that the structure roughly tracks the structure
of the precision matrix itself, and that the corrections are at the
1\% level. On the right we plot the diagonal entries for the entire
$\widetilde{D}$ matrix, and observe weak dependence of the corrections
on $r_{\parallel}$. }
\end{figure}

We begin by computing the $\widetilde{D}$ introduced above. In figure
\ref{fig:D} we show the result, with the primary corrections of $\approx1\%$.
Although the $\widetilde{D}$ we find is \emph{not} proportional to
the precision matrix, we find positive corrections on the diagonal,
and smaller negative corrections for adjacent bins, reflecting the
general structure of the precision matrix. 

In order to verify that the correction introduced above actually yields
a better inverse, we perform the following experiment. First, we split
the $n=10$ independent estimates $\hat{C}_{i}$ into five that will
be used to estimate the covariance matrix, and five that will be used
to generate an independent estimate of the precision matrix (by inverting
the covariance matrix). We multiply these together to get a matrix
close to the identity. To make the results more readable we present
$\widetilde{R}$, which is the average of this residual matrix over
all possible splits of the 10 estimates $\hat{C}_{i}$ into two groups
of five. Since our estimate of the covariance matrix is unbiased,
systematic deviations of $\widetilde{R}$ from the identity matrix
result from bias in the precision matrix, introduced by inversion. 

\begin{figure}
\includegraphics{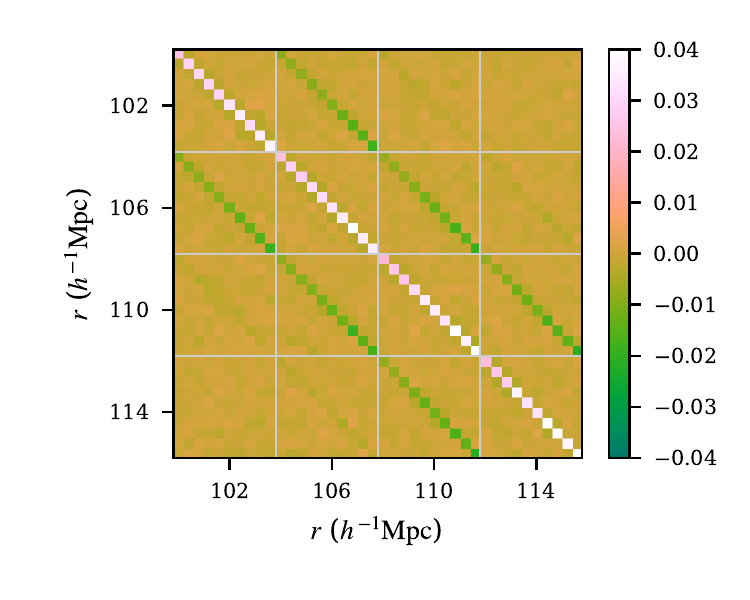}\hfill{}\includegraphics{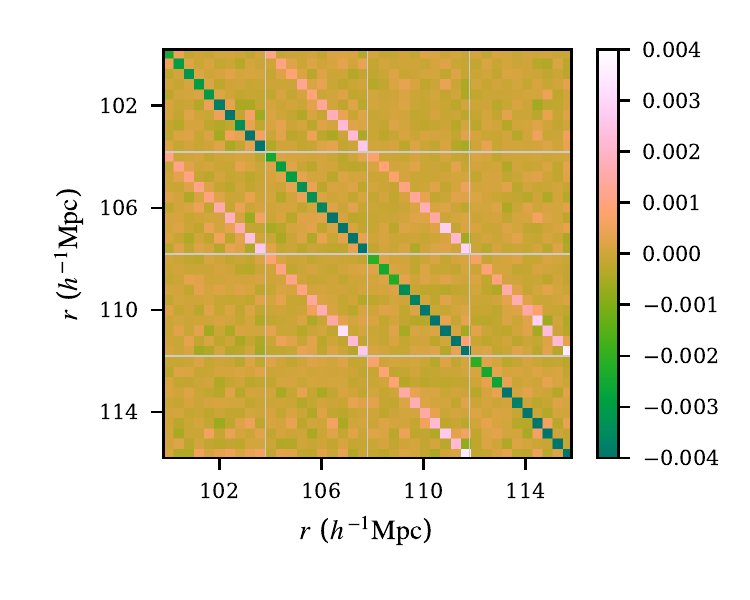}

\caption{\label{fig:R}A plot of $\tilde{R}-1$, the residual matrix described
above, both without (left) and with (right) the quadratic correction
to the precision matrix. \emph{Without} the correction we find residuals
of roughly 1\%, as we would anticipate based on the $\widetilde{D}$
we found. \emph{With} the correction we find that the residuals are
reduced by an order of magnitude but not eliminated entirely. }
\end{figure}

In figure \ref{fig:R} we plot a portion of $\widetilde{R}-1$ \emph{without
}the quadratic correction above and then \emph{with} the quadratic
correction applied. Without the quadratic correction we find systematic
deviations of $\widetilde{R}$ from the identity at around 1\%, as
we would anticipate from the $\widetilde{D}$ computed above. With
the quadratic correction we find that the residuals are reduced by
an order of magnitude, but not eliminated entirely. In other words
the quadratic correction reduces the bias in the precision matrix,
but does not provide an unbiased estimate. 

\subsection{Consistency of Fitting}

We now perform an alternative check on the inverse of the model covariance
matrix. In the main body of the paper we introduce a one-parameter
family of model covariance matrices, $C\left(a\right)$, then fit
the model against a sample covariance matrix, $\hat{C}$, computed
using 900 QPM mocks. The shot-noise rescaling parameter $a$ can be
estimated using two different likelihoods,
\begin{align}
-\log\mathcal{L}_{1}\left(a\right) & =\text{tr}\left[\Psi\left(a\right)\hat{C}\right]-\log\det\Psi\left(a\right).\\
-\log\mathcal{L}_{2}\left(a\right) & =\text{tr}\left[\hat{\Psi}C\left(a\right)\right]-\log\det C\left(a\right).
\end{align}
If we have accurate estimates of $\hat{\Psi}$ and $\Psi\left(a\right)$,
the two approaches should lead to consistent estimates of $a$. In
this subsection we will show that this consistency is achieved if
we apply the quadratic correction to $\Psi\left(a\right)$. 

We consider three different estimates of $a$:
\begin{itemize}
\item We will first estimate $a$ using the $\mathcal{L}_{2}$ likelihood.
Here we know that $\hat{\Psi}=\left(1-D\right)\hat{C}^{-1}$ provides
an unbiased estimate of $C^{-1}$, so this approach should yield an
unbiased estimate of $a$.
\item We then estimate $a$ using the $\mathcal{L}_{1}$ likelihood, with
$\Psi\left(a\right)$ the \emph{uncorrected} inverse of $C\left(a\right)$. 
\item Finally we estimate $a$ using the $\mathcal{L}_{1}$ likelihood,
with the quadratic correction applied to $\Psi\left(a\right)$. 
\end{itemize}
The fitting results are presented in table \ref{tab:a12}. In addition
to the central values we make jackknife estimates (leaving out individual
mocks) of the uncertainty in $a$. 
\begin{table}
\begin{centering}
\begin{tabular}{|l|c|}
\hline 
Method & $a$\tabularnewline
\hline 
\hline 
$\mathcal{L}_{2}$ likelihood & $1.0572\pm0.0038$\tabularnewline
\hline 
$\mathcal{L}_{1}$ likelihood, no correction & $1.0680\pm0.0015$\tabularnewline
\hline 
$\mathcal{L}_{1}$ likelihood, quadratic correction & $1.0590\pm0.0016$\tabularnewline
\hline 
\end{tabular}
\par\end{centering}
\caption{\label{tab:a12}Fitting results for the parameter $a$. The $\mathcal{L}_{2}$
likelihood should yield an unbiased result, while the $\mathcal{L}_{1}$
likelihood is sensitive to bias in $\Psi\left(a\right)$. In this
application, the quadratic correction is sufficient to bring the $\mathcal{L}_{1}$
estimate into agreement with the $\mathcal{L}_{2}$ estimate. We include
$1\sigma$ jackknife uncertainties.}
\end{table}

The fitting results in table \ref{tab:a12} suggest a significant
bias in $a$ when we use the \emph{uncorrected $\Psi\left(a\right)$
}for $\mathcal{L}_{1}$ fitting, but consistent results for $a$ when
we apply the quadratic correction to $\Psi\left(a\right)$. We note,
however, that when testing to see if two sets of estimates for $a$
are consistent with one another it is \emph{not }appropriate to simply
add the uncertainties in each estimate in quadrature, as this ignores
the possible issue of correlated errors in the $\mathcal{L}_{1}$
and $\mathcal{L}_{2}$ fittings. In order to address this we perform
a second test on $\Delta a$, the difference between $a$ as estimated
using the $\mathcal{L}_{1}$ likelihood and $a$ as estimated using
the $\mathcal{L}_{2}$ likelihood, with jackknife error bars computed
for $\Delta a$. The results, shown in table \ref{tab:Delta-a}, clearly
establish that in this application, the quadratic correction leads
to consistent fitting results. 
\begin{table}
\begin{centering}
\begin{tabular}{|c|c|}
\hline 
Correction to $\Psi\left(a\right)$ & $\Delta a$\tabularnewline
\hline 
\hline 
None & $0.0110\pm0.0035$\tabularnewline
\hline 
Quadratic & $0.0019\pm0.0035$\tabularnewline
\hline 
\end{tabular}
\par\end{centering}
\caption{\label{tab:Delta-a}Results for $\Delta a$, the difference between
$a$ estimated with the $\mathcal{L}_{1}$ likelihood and $a$ estimated
with the $\mathcal{L}_{2}$ likelihood. The quadratic correction leads
to consistent estimates of $a$ from the two likelihoods. We include
$1\sigma$ jackknife uncertainties.}
\end{table}

\end{document}